%% file: 00-main.tex
\documentclass[10pt,conference]{IEEEtran}
\IEEEoverridecommandlockouts
\usepackage{cite}
\usepackage{amsmath,amssymb,amsfonts}
\usepackage{algorithmic}
\usepackage{graphicx}
\usepackage{textcomp}
\usepackage{xcolor}

\definecolor{verylightgray}{rgb}{.97,.97,.97}
\usepackage{enumitem}

\usepackage{balance}
\usepackage{listings, xcolor}
\usepackage{pifont}
\usepackage{ulem}
\usepackage{array,framed}
\usepackage{setspace}
\usepackage{latexsym,fancyhdr,url}

\usepackage{enumerate}
\newcommand{\para}[1]{\vspace{2pt}\noindent\textbf{#1.~}}
\usepackage{xspace}
\usepackage{multirow}
\usepackage{microtype}
\usepackage{tcolorbox}
\usepackage{multirow}
\usepackage{colortbl}
\usepackage{diagbox}
\usepackage{tabularray}
\definecolor{Silver}{rgb}{0.752,0.752,0.752}

\def\BibTeX{{\rm B\kern-.05em{\sc i\kern-.025em b}\kern-.08em
    T\kern-.1667em\lower.7ex\hbox{E}\kern-.125emX}}
\begin{document}

\title{An Empirical Study on Embodied Artificial Intelligence Robot (EAIR) Software Bugs\\
%{\footnotesize \textsuperscript{*}Note: Sub-titles are not captured in Xplore andshould not be used}
}

\author{\IEEEauthorblockN{1\textsuperscript{st} Zeqin Liao}
\IEEEauthorblockA{\textit{Sun Yat-sen University} \\
%\textit{Sun Yat-sen University}\\
ZhuHai, China \\
liaozq8@outlook.com}
\and
\IEEEauthorblockN{2\textsuperscript{nd} Zibin Zheng}
\IEEEauthorblockA{\textit{Sun Yat-sen University} \\
%\textit{Sun Yat-sen University}\\
ZhuHai, China \\
zhzibin@mail.sysu.edu.cn}
\and
\IEEEauthorblockN{3\textsuperscript{rd} Peifan Reng}
\IEEEauthorblockA{\textit{Sun Yat-sen University} \\
%\textit{Sun Yat-sen University}\\
ZhuHai, China \\
rushfinen@gmail.com}
\and
\IEEEauthorblockN{4\textsuperscript{th} Henglong Liang}
\IEEEauthorblockA{\textit{Sun Yat-sen University} \\
%\textit{Sun Yat-sen University}\\
GuangZhou, China \\
lianghlong@mail2.sysu.edu.cn}
\and
\IEEEauthorblockN{5\textsuperscript{th} Zixu Gao}
\IEEEauthorblockA{\textit{Sun Yat-sen University} \\
%\textit{Sun Yat-sen University}\\
ZhuHai, China \\
gaozx9@mail2.sysu.edu.cn}
\and
\IEEEauthorblockN{6\textsuperscript{th} Zhixiang Chen}
\IEEEauthorblockA{\textit{Sun Yat-sen University} \\
%\textit{Sun Yat-sen University}\\
ZhuHai, China \\
chenzhx69@mail2.sysu.edu.cn}
\and
\IEEEauthorblockN{7\textsuperscript{th} Wei Li}
\IEEEauthorblockA{\textit{Sun Yat-sen University} \\
%\textit{Sun Yat-sen University}\\
ZhuHai, China \\
liwei378@mail2.sysu.edu.cn}
\and
\IEEEauthorblockN{8\textsuperscript{th} Yuhong Nan}
\IEEEauthorblockA{\textit{Sun Yat-sen University} \\
%\textit{Sun Yat-sen University}\\
ZhuHai, China \\
nanyh@mail.sysu.edu.cn}
}

\maketitle

\begin{abstract}

Embodied Artificial Intelligence Robots (EAIR) is an emerging and rapidly evolving technological domain. Ensuring their program correctness is fundamental to their successful deployment. However, a general and in-depth understanding of EAIR system bugs remains lacking, which hinders the development of practices and techniques to tackle EAIR system bugs.

To bridge this gap, we conducted the first systematic study of 885 EAIR system bugs collected from 80 EAIR system projects to investigate their symptoms, underlying causes, and module distribution. Our analysis takes considerable effort, which classifies these bugs into 18 underlying causes and 15 distinct symptoms, and identifies 13 affected modules. It reveals several new interesting findings and implications which help shed light on future research on tackling or repairing EAIR system bugs.
Firstly, among the 15 identified symptoms, our findings highlight 8 symptoms specific to EAIR systems, which is characterized by severe functional failures and potential physical hazards. Second, within the 18 underlying causes, we define 8 EAIR-specific causes, the majority of which stem from the intricate issues of AI-agent reasoning and decision making. Finally, to facilitate precise and efficient bug prediction, detection, and repair, we constructed a mapping between underlying causes and the modules in which they most frequently occur, which enables researchers to focus diagnostic efforts on the modules most susceptible to specific bug types.

\end{abstract}


\input{01_body}

{
     \bibliographystyle{ieeetr}
    \balance
    \bibliography{bib}
}

\end{document}

%% file: 01_body.tex
\section{Introduction}
\label{introduction}

%具身智能关节机器人（EAIHR）作为新兴技术，相关的大量项目已经开始在公众环境下快速产生，例如特斯拉Optimus[xx]，Figure 01 [xx]，Unitree H1。然而，EAIHR技术的可靠性和安全性高度依赖软件系统，而软件系统固有的缺陷（如软件Bug）可能引发严重的安全问题，甚至可能对用户的健康和财产构成重大威胁 [xx]。因此，具身智能关节机器人软件中的缺陷（bug）属于安全和可靠性关键型问题（Safety-critical）。如何有效避免或检测此类软件缺陷，已成为具身智能关节机器人领域重要且亟待解决的研究课题。
%\ZQ{todo: clear definition}

Embodied Artificial Intelligence Robots (EAIR), as a newly emerging technology, have begun gradually available in the public, such as Tesla Optimus~\cite{Optimus}, Nvidia Isaac-GR00T N1~\cite{GROOT} and Unitree H1~\cite{Unitree}. 
In contrast to traditional robot (e.g., ROS-based drones and autonomous driving), EAIR employs embodied agents (e.g., AI agent) to control their physical body to support four core-mechanisms, which includes: 1) embodied interaction such as reasoning-based question answering and grasping; 2) embodied perception such as reasoning-based sensing and real-time exploration, 3) embodied control such as behavior learning from surrounding interaction instead of predefined rules or well-trained models; and 4) embodied simulation such as fine-grained object interaction and multi-agent compatibility (see Section~\ref{EAIRs})~\cite{liu2024aligning, xu2024survey, roy2021machine}.
However, the reliability and security of EAIR technology heavily rely on systems software , whose inherent defects (i.e., bugs) can cause severe security issues and may even pose significant threats to users’ health and property~\cite{liu2024aligning}. Hence, ensuring the program correctness of an EAIR system software is crucial for its success~\cite{xu2024survey}.

%尽管如此，在EAIR系统中有效检测bug仍然是一项巨大的挑战[48, 64]。其中一个重要问题是，我们至今缺乏对EAIR bug的一般性和深入性的理解。例如，我们尚不清楚EAIR bug是如何被引入的（即根本原因），这些缺陷具体如何表现（即根本诱因所导致的现象），这些bug具体如何分布（即bug分布），以及如何深入揭示这些缺陷产生的识别机制（即检测机制）。

%此前已有多项研究开展了不同机器人领域中软件bug的分析，他们大致可以分为三类，（1）robotic vehicles （如基于ROS的drone）[xx]；（2）industrial control system (如工业机械比) [xx] 和 自动驾驶 [xx]。由于前两种是传统的rule-based 的机器人系统，不涉及与AI agent 的交互，他们的bug 分析结果不能直接用于EAIR的分析。尽管自动驾驶也是一种基于AI的机器人系统，但是由于它们与EAIR截然不同的系统架构，其分析结果也无法在EAIR系统中适用。举例，区别于无人驾驶，EAIR拥有四种特有的机制，AI agents, embodied interaction, multi-modal information aggregation, and sim-to-real adaptation，自动驾驶工作无法适用于解决这些机制相关的bug （见section 2.1）。 
%目前尚未有任何研究专门针对具身智能关节机器人（EAIR）中的bug特征进行系统化分析，更不必说实现自动化的bug检测工具了。

One important gap is the current lack of a general and in-depth understanding of EAIR bugs~\cite{duan2022survey}. For example, it is still unclear how these bugs are introduced (i.e., underlying causes), how they manifest (i.e., adverse symptoms), how is the relationship between underlying causes and symptoms, and how they are distributed (i.e., bug distribution).
To the best of our knowledge, the most relevant research on EAIR system analysis is a limited number of survey studies~\cite{duan2022survey, liu2024aligning,roy2021machine,xu2024survey}.
None of these studies can support a systematic examination of EAIR system bugs. As a result, it is difficult to developing detection techniques and distill lessons on tackling EAIR system bug.

\para{Our work} In this paper, we present the first systematic empirical study on the bug characteristics for the EAIR system, to the best of our knowledge. %In this capability, our empirical findings can facilitate the avoidance and diagnosis of EAIR system bugs during development and support their detection and repair after deployment.

%我们的发现可以帮助在开发过程中去避免EAIR system bug，在部署后去检测和修复 EAIR system bug
To this end, we manually construct the EAIR bug dataset, which can be divided into two steps. Firstly, we collected 80 EAIR system projects from industry, academic, and open-source platforms (e.g., GitHub~\cite{Github}). To the best of our knowledge, our EAIR system dataset is the most comprehensive dataset available in the public (see Section~\ref{EAIRSDataset}). 
Secondly, by leveraging 6,813 issues/PRs and 17,127 commit records drawn from the repositories of 80 EAIR projects, this study identifies and analyzes 885 bugs pertinent to EAIR systems. To the best of our knowledge, this represents the first and most comprehensive dataset of EAIR bugs (see Section~\ref{EAIRBugDataset}).

%\item What are the common patterns for detecting these EAIR system bugs? How these bugs can be found?
%回答这些研究问题对于开发者和研究人员均具有重要的实际意义。RQ1 归纳了常见的故障类型（见Section4），RQ2研究了缺陷的不良表现症状（见Section5），可帮助开发者在应用开发的早期阶段规避EAIR缺陷。 至 RQ3则总结了bug的分布规律（见Section6）， RQ4则研究了bug的通用patterns及其检测方法（见Section7），为研究人员设计effective and efficiency的缺陷检测技术提供指导。此外，通过分析 RQ1 至 RQ4 结果之间的相关性，我们能够获得更深入的见解。

Further, we investigate the research questions (RQs):

\begin{itemize} []
    \item \textit{RQ1 (Symptoms):} What are the adverse impact of these EAIR system bugs?
    \item \textit{RQ2 (Underlying causes):} What are the underlying causes of these EAIR system bugs? 
    \item \textit{RQ3 (Module distribution):} What is the bug distribution on different modules of the EAIR system? 
\end{itemize}

Guided by these RQs, a rigorous empirical analysis on 885 bug and their associated commit metadata reveals (i) 15 kinds of observable symptom, (ii) 18 distinct underlying causes, and (iii) 13 EAIR software module that demonstrate pronounced bug concentration in real-world settings. The inter-relationships among these three dimensions are subsequently analyzed. 

%我们从 80个EAIR系统项目的代码库的 6813 issues/PRs and 17127 commits中，研究了共计 885 个EAIR系统相关的缺陷（bugs）。通过对这些缺陷及其提交信息的人工分析，我们共识别出 18 种原因（underlying causes）、15 种缺陷表现症状（symptoms），以及 13 类EAIR软件模块，这些组件在实践中表现出显著的缺陷集中趋势。我们进一步评估了这三类现象之间的关系。在此基础上，本文提出了针对EAIR系统的软件测试、分析与修复的若干未来研究方向，以期为该领域的可靠性保障提供指导。

%Answering these research questions holds significant practical value for both developers and researchers. Specifically, RQ1 identifies common bug types (see Section~\ref{EvaluationtoRQ1}), while RQ2 reveals the adverse impacts of bugs (see Section~\ref{EvaluationtoRQ2}) which can help developers avoid EAIR-related bugs early in the development. RQ3 reveals the complexity of these bugs (see Section~\ref{Cause and Symptoms}), and RQ4 analyzes the distribution patterns of these bugs (see Section~\ref{EvaluationtoRQ3}). 

Through this study, we discover a set of novel and intriguing findings:
(1) Our investigation uncovered eight EAIR-specific symptoms. unlike other symptoms, EAIR-specific symptoms are associated with severe functional failures and physical hazards (Finding 1). The manifestation of EAIR-specific symptoms indicates that inherent challenges(e.g., AI hallucination, EAIR behavior learning, and sim-to-real logic) remain insufficiently addressed by current developers.
%在症状方面，我们发现了8种EAIR-specific 症状，区别于其他症状，这些症状会带来严重的功能性危害。同时这些症状的出现也暗示了，像AI幻觉、EAIR的行为学习、仿真到现实的逻辑这些固有的挑战没有被开发者很好地解决，
%
(2) Most of EAIR bug are the errors specific to EAIR core-mechanisms (i.e., embodied perception, embodied interaction, embodied control, and embodied simulation) (Finding 3). We also define eight EAIR-specific underlying causes, the majority of which stem from the intricate issues of AI-agent reasoning and decision making (Finding 4). Particularly, many of these EAIR-specific causes give rise to compound bugs, which further complicating their identification (Finding 7).
%
%在underlying cause 方面，大多数的EAIR bug都是EAIR核心机制（xxx）的错误。我们还定义了8中EAIR-specific cause，其中大部分都是AI agent相关的复杂的问题。更严重的是，大部分EAIR-specific cause 都会导致复合的bug，这进一步增加了识别的难度。
%
(3) We reveal a correlation relationship between EAIR-specific symptoms and EAIR-specific underlying causes, that is, bug arising from EAIR-specific causes are likely to manifest as high-severity EAIR-specific symptoms (Finding 6).
%
%在分析了症状与underlying cause之间的关系后，我们发现EAIR-specific cause 和EAIR-specific symptom 是紧密相关的，即EAIR-specific cause容易造成高危害性的EAIR-specific symptom。
%
(4) To facilitate precise and efficient bug prediction, detection, and repair, we constructed a mapping between underlying causes and the modules in which they most frequently occur (Table~\ref{causeandmodule}). This mapping enables researchers to focus diagnostic efforts on the modules most susceptible to particular bug types. Moreover, we discover that invocation bugs in EAIR systems constitute a promising research frontier warranting deeper exploration (Finding 9).
%%在分布方面，我们建立了underlying cause与模块之间的映射关系，将来研究人员可以根据bug的类型定位相关的模块，用于实现准确高效的bug prediction, detection and repair。此外，我们还发现，Invocation bugs in EAIR systems represent animportant research direction that merits deeper exploration in future studies。
Further, we also established a mapping between symptoms and system modules (Table~\ref{symptomandmodule}). Practitioners can leverage this information to allocate more resources(e.g., development time and effort) toward modules which possess the most severe errors or the greatest latent risk.

In summary, this paper makes the following contributions:

\begin{itemize}
    \item  To the best of our knowledge, we conduct the first systematic study to investigate bugs in embodied artificial intelligence robot systems.
    \item We study the EAIR bugs from different perspectives (underlying causes, bug symptoms, and bug distribution, and summarize EAIR-specific findings.
    \item We enumerate the implications of our findings which help shed light on tackling EAIR bugs, from the perspectives of developers and researchers.
    %\item  We construct the first dataset of the EAIR system as well as the first dataset of 885 EAIR bugs. We also establish a reproducible platform consisting of 80 Dockers for the EAIR system dataset. 
    \item We will release our empirical study results as well as the corresponding datasets~\footnote{\url{https://doi.org/10.6084/m9.figshare.28631276.v1}}.All of these can serve as the basis of our study and future research in this direction.
\end{itemize}

\section{Background}
\label{EAIRs}

%\para{Embodied artificial intelligence and embodied agent}
%Embodied Artificial Intelligence (EAI) refers to the paradigm wherein artificial agents learn from the interactions between them and surrounding environments. 
Embodied agent is the most essential element of Embodied Artificial Intelligence~\cite{duan2022survey}. Prior studies~\cite{liu2024aligning, xu2024survey, roy2021machine} indicate that typical embodied agent encompass deep-learning models, reinforcement-learning, large language models, and world models.
%An embodied agent generally exhibits two core competencies: 1) high-level embodied task planning (i.e., the decomposition of abstract and complex objectives into concrete sub-tasks) and 2) low-level embodied action planning(i.e., the incremental execution of those sub-tasks by leveraging the policy capabilities of foundational models).
With the excellent task-decomposition mechanism, embodied agents enable robots to perform a wide range of complex operations in real-world settings such as embodied perception, embodied interaction, and embodied control~\cite{xu2024survey}.
%具身智能（EAI）指的是，让人工智能体在受到物理约束下，通过与周围环境的交互进行学习。
% 具身智能体是具身人工智能最重要的基础。早先工作表明，常见的具身智能体包括深度学习、强化学习，大语言模型和世界模型。

%具身智能体通常具备以下能力：1）将抽象复杂的任务分解为具体的子任务，这被称为高级具身任务规划。2）通过有效利用基础模型的策略功能，逐步实现这些子任务，这被称为低级具身行动规划。凭借这种优秀的任务分解规划能力，具身智能体可以使能机器人在信息丰富且复杂的现实世界开展各种复杂任务，包括具身感知，具身交互，具身控制。

\begin{figure}[t]
\centering
\includegraphics[width=3.5in]{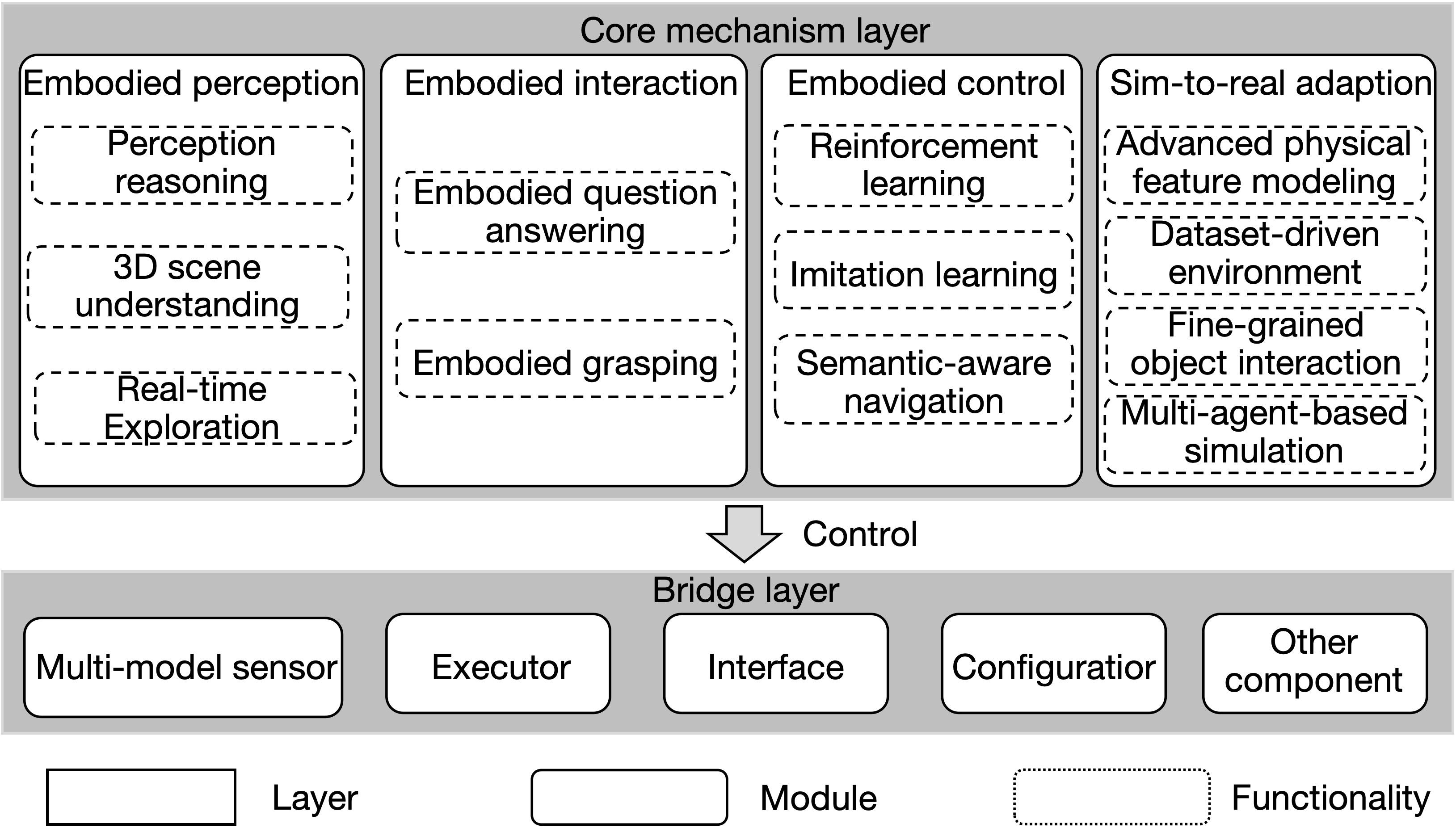}
\caption{ The embodied artifical intelligent robots system architecture.} 
\label{EAIR_architecture}
\vspace{-2mm}
\end{figure}

\para{Embodied artificial intelligence robot}
Embodied artificial intelligent robots (EAIR) refer to robots that employ embodied agents to control their physical body to conduct interaction and learning with their surrounding environment~\cite{duan2022survey}, for instance, Tesla Optimus~\cite{Optimus}, Figure 02~\cite{figure}, and the Unitree H1~\cite{Unitree}. As illustrated in Figure~\ref{EAIR_architecture}, a typical EAIR system architecture can be divided into two layers, including the core-mechanism layer and the bridge layer.

%具身智能机器人是一种利用具身智能体来控制物理身体来实现周围环境进行交互和学习的机器人。如图xxx所示，一个典型的EAIR系统可以分为两层：核心机制层和桥接层。

%\ZQ{todo: narrow down}

\para{Core mechanism layer} 
The core-mechanism layer constitutes the principal discriminator between EAIR system and traditional robotic system. Typically, EAIR systems contain a part or all of the following four modules which are enabled by the embodied agents~\cite{liu2024aligning, xu2024survey, roy2021machine}.
%核心机制层是区分具身智能机器人与传统机器人的关键。核心机制层通常包括由具身智能体赋能的四个模块中的部分或全部。

\begin{itemize} [leftmargin=10pt] 

\item \textbf{Embodied interaction.}
%The embodied interaction module enables the agent to interact with humans and the environment in physical or simulation environment.
A typical embodied interaction module possess the functionality including embodied question answering (EQA) and embodied grasping~\cite{liu2024aligning}.
For EQA, the embodied-interaction module is required to explore the environment from an egocentric perspective, for gathering the evidence necessary to resolve the query. The agent within this module must reason about which exploratory actions to undertake and the appropriate juncture at which exploration should terminate and an answer should be returned.
Beyond information gathering, the module must also execute human-issued commands such as grasping or placing objects, thereby mediating interaction among the robot, the human, and the surrounding objects. Achieving reliable embodied grasping demands comprehensive semantic understanding, scene perception, decision-making, and control.

%具身交互模块旨在实现智能体在物理或模拟空间中与人类及环境进行交互。典型的具身交互模块的功能包括包括具身问答（EQA）和具身抓取，这是以往机器人系统所不具备的。对于 EQA 任务，具身交互模块需要从第一人称视角探索环境，以收集回答给定问题所需的信息具身交互模块的代理不仅必须考虑采取哪些行动来探索环境，还必须确定何时停止探索以回答问题。
%此外，具身交互模块还涉及基于人类指令执行操作，例如抓取和放置物体，从而完成机器人、人类和物体之间的交互。具身抓取需要全面的语义理解、场景感知、决策和鲁棒控制规划。

\item \textbf{Embodied perception.}
Unlike traditional robotic that rely on low-level and two-dimensional map of sparse or semi-dense grids (e.g., vacuum cleaner robot),
embodied-perception requires a substantially richer understanding of object relationships and dynamic surroundings~\cite{xu2024survey}. 
%This module integrates the functionality perception reasoning, 3D scene understanding and active exploration.
For perception reasoning, the module must detect, interpret, and reason the semantic roles and relationships of objects within the environment, and this reasoning capability is largely absent from earlier robotic systems. 
In addition, it must locate these objects in three-dimensional scene data and estimate their geometric attributes.
Traditional robots engage in passive perception (i.e., constructing maps only when prompted by external instructions),
and therefore struggling with continuously evolving real-world settings. embodied-perception module should possess the ability of real-time and active exploration 
on the surroundings.

%区别于传统的依赖于2D低级地图（例如稀疏地图、半密集地图）机器人系统（如扫地机器人），具身感知模块能够对空间中的物体和动态环境有更深入的理解。具身感知模块具有感知和推理，理解场景物体中的三维关系，和实时主动探索等功能。
%对于感知和推理，具身感知模块需要对场景中的物体的语义进行识别、解释和推理，这种推理能力是以往机器人系统所不具备的。此外，具身感知模块还涉及3D场景数据中识别其位置并推断其几何属性。
%传统机器人是被动感知环境，环境地图构建依赖于外部指令，因此无法适应不断变化的场景。由于机器人能够运动并与周围环境频繁交互，具身感知模块应该具备实时主动探索和感知环境的能力。

%并基于视觉信息预测和执行复杂的任务。

% 具有具身感知的智能体必须在物理世界中移动并与环境交互。这需要对三维空间和动态环境有更深入的理解。具身感知需要视觉感知和推理能力，理解场景中的三维关系，并基于视觉信息预测和执行复杂的任务。

%主动视觉感知系统需要状态估计、场景感知和环境探索等基本能力
%传统的 vSLAM 系统利用图像信息和多视图几何原理来估计机器人在未知环境中的姿态，从而构建由点云组成的低级地图（例如稀疏地图、半密集地图和密集地图）由于低级地图中的点云并不直接对应于环境中的物体，因此具身机器人难以对其进行解释和利用。语义概念的出现，尤其是集成了语义信息解决方案的语义vSLAM系统的出现，显著提高了机器人感知和导航未知环境的能力。

%3D场景理解旨在从3D场景数据中区分物体的语义、识别其位置并推断其几何属性，这是自动驾驶的基础。
%与图像不同，点云稀疏、无序且不规则，[120]使得场景解读极具挑战性。

%前面介绍的3D场景理解方法赋予机器人被动感知环境的能力，这种情况下感知系统的信息获取和决策能力无法适应不断变化的场景。然而，被动感知是主动探索的重要基础。机器人既然能够运动并与周围环境频繁交互，就应该具备主动探索和感知环境的能力。它们之间的关系如图7所示。目前解决主动感知问题的方法主要集中在与环境的交互上。112，113]或通过改变观看方向来获取更多视觉信息[114，115，116，117]。

%与在平面图像范围内运作的传统 2D 视觉接地 (VG) 不同，3D VG 融合了物体之间的深度、透视和空间关系，为智能体与环境交互提供了更稳健的框架。3D VG 的任务是使用自然语言描述在 3D 环境中定位物体。

%视觉语言导航 (VLN) 是具身人工智能 (Embodied AI) 的一个关键研究问题，旨在使代理能够按照语言指令在未见过的环境中导航。VLN 要求机器人理解复杂多样的视觉观察，并同时解释不同粒度的指令。VLN 的输入通常包含两部分：视觉信息和自然语言指令。视觉信息可以是过去轨迹的视频，也可以是一组历史-当前观察图像。自然语言指令包括具身代理需要达到的目标或预期完成的任务。具身代理必须使用上述信息从候选列表中选择一个或一系列动作来满足自然语言指令的要求。

%前面介绍的3D场景理解方法赋予机器人被动感知环境的能力，这种情况下感知系统的信息获取和决策能力无法适应不断变化的场景。然而，被动感知是主动探索的重要基础。机器人既然能够运动并与周围环境频繁交互，就应该具备主动探索和感知环境的能力。它们之间的关系如图7所示。目前解决主动感知问题的方法主要集中在与环境的交互上。112，113]或通过改变观看方向来获取更多视觉信息[114，115，116，117]。

\item \textbf{Embodied simulation.}
In contrast to traditional robot (i.e., autonomous driving), EAIR systems demand more fine-grained simulation that approximate the richness and dynamics of the real world~\cite{duan2022survey}. 
%Embodied simulation requires to support: 1) advanced physical-feature modelling, 2) dataset-driven scene construction, 3) fine-grained object interaction, and 4) multi-agent compatibility.
First, beyond the basic physical feature like collisions, rigid-body dynamics and gravity, embodied simulation must faithfully model advanced physical-feature including cloth, fluid dynamics and soft-body physics. 
Second, whereas traditional simulation typically rely on handcrafted asset libraries, embodied simulation module further leverages large-scale datasets to generate diverse and dynamically evolving environments. 
Third, embodied simulation extend beyond the basic interactivity found in traditional robot (e.g., collision handling) by enabling fine-grained interactivity (e.g., multi-state transitions). For example, the change of an apple to discrete slices after cutting.
Finally, the embodied simulation require to support both cooperative and adversarial multi-agent scenarios.
%相较于传统机器人系统（如自动驾驶），具身智能机器人需要构建更复杂的更接近于真实环境的要求的仿真环境。具身仿真具有高级物理特征建模、数据集驱动的环境构建、细粒度对象交互以及多智能体兼容等功能。
%区别于传统机器人，具身仿真不仅需要包括碰撞、刚体动力学和重力等基本物理特征建模，还需要对包括布料、流体和软体物理等高级物理特征进行建模
%与传统机器人主要依赖于资产构建环境不同的是，具身仿真通过数据集驱动的环境构建实现了更全面动态的建模。
%具身仿真不仅支持传统机器人仿真的基本交互性，例如碰撞；还支持对象更细粒度的交互性，例如多状态变化。例如，当苹果被切片时，它的状态会变为苹果片。
% 具身仿真需要实现多智能体兼容，实现对抗式和协作式智能体的仿真。

%高级物理特征建模：将其分为基本物理特征（B）和高级物理特征（A）。参考图2，基本物理特征包括碰撞、刚体动力学和重力建模，而高级物理特征包括布料、流体和软体物理。

%数据集驱动的环境：第一种类型是数据集驱动环境;第二种类型是资产驱动环境，基于游戏的具身AI模拟器更有可能从资产商店获取对象数据集，而基于世界的模拟器则倾向于从现有的3D对象数据集中导入对象数据集。

%细粒度对象交互：一些模拟器仅支持对象的基本交互性，例如碰撞。高级模拟器则支持对象更细粒度的交互性，例如多状态变化。例如，当苹果被切片时，它的状态会变为苹果片。因此，我们将这些不同级别的对象交互分为可交互对象模拟器（I）和多状态对象模拟器（M）。

%虚拟现实控制器：虚拟现实控制器接口提供更具沉浸感的人机交互，并方便使用现实世界的对应设备进行部署。

%导航（N）、原子动作（A）和人机交互（H）。导航是最低层，是所有具身人工智能模拟器的共同特征[38]。机交互类别，因为它们旨在提供高度逼真的基于物理的模拟和多种状态变化。这只有通过人机交互才有可能，因为在与这些虚拟对象交互时需要人类水平的灵活性。

% ：模拟器需要具备丰富的对象内容，才能构建用于对抗式和协作式人工智能体训练[39] [40]的多智能体特征，并使其具有实际价值。由于缺乏支持多智能体的模拟器，利用这些具身化人工智能模拟器中的多智能体特征的研究任务也较少

\item \textbf{Embodied control.}
Unlike traditional robots that rely on predefined rules or well-trained models, embodied-control module learns from the interaction with its surroundings, and refines EAIR control behaviour via reward-based method~\cite{roy2021machine}.
%Embodied-control module integrates the functionalities of reinforcement learning (RL), imitation learning (IL), and semantics-aware navigation. 
By leveraging RL and IL paradigms, the module can adapt to dynamic real-world environment instead of specific scenes. Moreover, instead of depending heavily on simple cues such as location (i.e., the paradigm of traditional robot), embodied control extend to reasons over inferred object semantics, their relational context, and their geometric attributes, yielding more reliable navigation performance.
\end{itemize}
%相较于传统机器人基于预定义的规则或训练好的模型，具身控制通过与环境的交互进行学习，并利用奖励机制优化行为以获得最优策略，从而避免了传统机器人自适应能力差的弊端。
%具身控制具有强化学习、模仿学习和基于语义的导航等功能。具身控制利用强化学习、模仿学习等技术从而探索未知领域，而非适应非结构化环境。区别于传统机器人，具身控制不仅利用位置信息，还利用利用推理的物品语义关联信息和几何属性来实现更可靠的导航效果

%具身控制方法可以分为两类：

%具身控制将强化学习 (RL) 与模拟到现实 (sim-to-real) 技术相结合，通过环境交互来优化策略，从而探索未知领域，超越人类能力，并适应非结构化环境。虽然机器人可以模仿许多人类行为，但有效完成任务通常需要基于环境反馈进行强化学习训练。最具挑战性的场景包括接触密集型任务，其中操作需要根据反馈（例如被操作物体的状态、变形、材质和受力）进行实时调整。在这种情况下，强化学习必不可少。

%具身智能关节机器人是指一种通过关节式的物理体（身体）与环境进行交互，并利用AI agent 进行智能决策的机器人。得益于着其智能决策能力和特有的关节形态，具身智能关节机器人具有适用于各种应用场景的巨大潜力，已经在真实世界开始投入商业使用，如特斯拉optimus [xxx]，Figure 01 [xx] and Unitree H1 [xx]. 具身智能关节机器人通常可以分为以下的几个重要的类型[xx]，包括 Humanoid Robots(如特斯拉optimus)，Quadruped Robots(Boston Dynamics Spot), 固定基座基座机器人 (Franka Emika Panda), 以及 Biomimetic Robots

\para{Bridge layer} This layer establishes the interaction between the system and hardware of EAIR according to the control instructions, which can be divided into the following modules:
\begin{itemize} [leftmargin=10pt] 
    \item \textbf{Multi-modal sensor.} This module is responsible for extracting useful information from sensor data.
    \item \textbf{Executior.} This module responds to control instructions and controls the robot to execute actions or perform movements.
    %该模块响应控制指令，控制机器人做出行为或移动
    \item \textbf{Interface.} This module provides various types of internal and external interfaces.
    %该模块提供各种形式的内部或外部接口
    \item \textbf{Configurator.} This module implements various complex configurations for EAIR.
    %该模块实现EAIR的各种复杂配置
\end{itemize}

\begin{figure*}[t]
\centering
\includegraphics[width=6.5in]{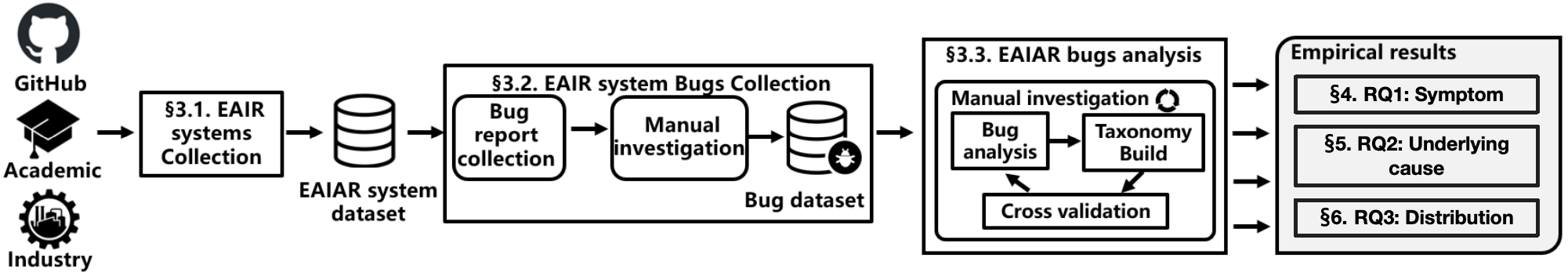}

\caption{ The overall analysis procedure of our research.} 
\label{overview}
\vspace{-2mm}
\end{figure*}

%\subsection{Bug-fixing Process in EAIR system}
%\label{EAIRs}

\section{Methodology}
\label{methodology}

\subsection{EAIR systems Collection}
\label{EAIRSDataset}

We aim to select the representative EAIR system projects as our research subjects. Specifically, by applying the following unique criteria, we collect EAIR system projects from the sources of industry, academic, and open-source platforms (i.e., GitHub).

\para{Criterion \#1 (Relevance)} The system projects that we collect must be closely related to our EAIR topic. To this end, we utilize a set of keywords such as "Embodied", “Artificial intelligence”, “Agent” and “Robot” to exhaustively search for system projects from related sources. %Within this step, we obtain xxx candidate system projects.

\para{Criterion \#2 (Availability)} The collected system projects should be public available. For example, we exclude the closed-source EAIR system projects such as Optimus in Tesla~\cite{Optimus}. Nonetheless, our dataset remains 9 mainstream commercial EAIR system projects including Unitree robot~\cite{Unitree}, 
Isaac-GR00T N1 in Nivdia~\cite{GROOT},
Habitat in FaceBook~\cite{Habitat}, and so on.

\para{Criterion \#3 (Real-world deployment)} 
%The system projects that we gather must be real-world deployable projects. Specifically,
To ensure the practical significance of our analysis, we only select the EAIR system projects that support deploying on real-world robot hardware and exclude the other 8 candidate system projects lacking support.

%此步骤的目的是为我们的研究选择具有代表性的EAIR应用程序对象。我们利用一下两个规则从工业界（商业EAIRs项目），学术论文，Github上收集一组候选的EAIRs应用程序
%Criterion #1 (Relevance)应用程序必须是密切相关的。我们使用关键字如Embodied, Artifical intelligence, Agent, Robot 来从先关来源穷尽搜索相关项目。我们收集到xxx个候选项目

%Criterion #2 (Availability):应用程序必须是可公开访问的。举例，在这一步我们排除了xxx个闭源的EAIRs软件（比如特斯拉optimus）。尽管如此，我们还是收集了xx个主流的业界商业EAIRs项目，比如宇树，facebook的habitat，xxx等

%Criterion #3 (Reality): 应用程序必须是真实世界中的EAIRs应用程序。我们只选择那些已经在真实机器人硬件上部署或有足够证据支持部署的EAIRs应用程序。我们排除xxx个不支持真实部署的项目

%Criterion #4 (Representativeness and Popularity) 为了选择具有代表性的应用程序，我们首先尽可能地保留了商业EAIRs项目，共计xxx个。而对于那些来源于学术圈和Github的项目，我们调查（1）他们收到的引用（#Citation）和(2)Github仓库的star数量（#Star）直到2025年3月。我们排除了xxx个#Citation<200和#Star<50的项目。

\para{Criterion \#4 (Representativeness and Popularity)} To select the representative system projects, we have prioritized retaining commercial EAIR projects to the extent possible, with a total of 9 system projects. For those projects that derive from academic and GitHub, we investigate (1) the citations that they receive (\#Citation) and (2) the stars of GitHub repositories (\#Star) until March 2025. We exclude 13 system projects within the conditions of $"\#Citation<200"$ and $"\#Star<50"$.

%根据上述unified criteria，我们构建了首个EAIR systems数据集，由xxx个真实和流行的EAIR systems项目组成。这一过程耗费了我们大量的人工努力（近八个人月），因为EAIR应用的穷尽搜索是极其耗时的，复现EAIR应用以检查它是否能部署是很困难。据此，我们为这xxx个EAIR systems项目封装了docker，形成一个可重现平台，which can serve as the basis of future research in the direction.

%\para{EAIR system dataset construction} 
Based on the above-stated criteria, we construct the EAIR system dataset with 80 real-world and popular EAIR system projects. This step consumes considerable manual effort (i.e., about 8 person-months) because it is extremely time-consuming to reproduce these projects to check whether they are deployable. In this way, we have constructed Dockers for these 80 EAIR system projects, forming a reproducible platform that can serve as the basis of future research in the direction.

\subsection{Bugs Collection}
\label{EAIRBugDataset}

\para{Bug report collection}
Considering that the EAIR systems of our dataset all have open-source repositories on GitHub, we first collect the EAIR system bug reports by applying GitHub API to extract two data contents, i.e., issues/pull requests(PRs) and commits. 
For each issue/PR, we collect information including headings, texts, comments, events, and bug labels. For each commit, we collect headings, commit information, corresponding files, and ID/URL.
%In addition, we also extract the code patches of certain commits filtered by Section~?????????????????????????????????????????????????????????\ref{Method}, for further bug pattern analysis in Section~?????????????????????????????????????????????????????????\ref{EvaluationtoRQ4}. 
As for January 2025, we finally obtained the original EAIR system bug reports with a total of 6813 issues/PRs and 17127 commits.

%我们的EAIR system Bug库是通过利用GitHub API，从两个数据来源构建的：问题（issues）和提交（commits）。对于问题，我们收集每个issue/PR的所有信息，包括问题标题、问题正文、评论、事件和漏洞类别标签。对于每个提交，我们收集其标题、提交信息、受影响的文件以及ID/URL；对于根据§3.3筛选后的某些提交，还会收集其实际的代码更改块，并根据§3.3进行处理，随后用于§7中的代码模式分析。到2025年1月，我们共收集了xxx个已关闭的issues/PR和xxx个提交作为原始数据集，详细见表2。

\para{Bug database construction}
This procedure aims to establish a valid EAIR system bug database rigorously. Firstly, we filter out the issues/PRs which have explicit bug labels (e.g., "Bug", "Issue") and receive the feedback of the developer (e.g., response, close). We pay more attention to issues/PRs with developer feedback because they commonly reflect sufficient bug analysis information. In this way, we narrowed down the EAIR system bug reports to 1184 issues/PRs and 950 commits.

%此步骤旨在严格构建一个有效的EAIR软件bug的数据集。首先，我们筛选出那些明确标记为bug（例如，“Bug”，“Issue”）并且收到开发者反馈（如回复、关闭等）的问题。我们重点关注开发者反馈的问题，因为它们已经被开发者处理，并通常包含足够的bug分析信息。通过这种方式，我们获得了xxx个漏洞报告。

Further, our researchers manually inspect all of these bug reports and delete the invalid bug reports such as non-reproducible reports, functional enhancement patches, and repetitive reports. Specifically, among the 1184 issues/PRs and 753 commits, we only retain the bug reports that satisfy two conditions. 

\para{Condition \#1 (Reproducibility)} We confirm that a given bug report contains a valid EAIR system bug, only if (1) the bug can be reproduced under the vulnerable version used in our research, or (2) the bug report contains the distinct videos or images which describe rigorous reproduction process. 

\para{Condition \#2 (Adverse Symptoms)} We determine a given bug report to be valid only if it has undesirable or harmful symptoms on the EAIR system.

Through this process, we finally obtained 885 valid bugs for forming the first EAIR system bug database, to the best of our knowledge. In our bug database, each EAIR system bug consists of the following parts: (1) a vulnerable version of the EAIR system, (2) videos or images that describe a detailed reproduction process, and (3) corresponding bug patches if there are any. Since it is significantly difficult to reproduce the EAIR system bug (e.g., complex environment setup), we spent a lot of manual effort with a total of 6 person-months in this process. 
%Further, we present the detailed information for our constructed bug database in column Table~?????????????????????????????????????????????????????????\ref{}.

%基于这xxx个漏洞报告，我们手动检查了每一个报告，排除了无效的报告（例如重复的、标签错误的、功能增强），最终得到了xxx个有效漏洞报告。
%具体来说，在xxx个功能性漏洞报告中，我们仅保留符合以下三个标准的漏洞报告。首先，相关的EAIR软件bug必须能够在我们研究时使用的有缺陷的应用版本上复现，或者即使我们未能自行复现该bug，漏洞报告中应包含清晰的bug重现视频/图片。
%其次，相关的bug必须与补丁明确关联，因为补丁有助于分析漏洞的根本原因。通过这种方式，我们获得了xxx个有效EAIRs漏洞。
%对于每个有效的EAIR软件漏洞，我们准备了以下内容：(1) 有缺陷的应用，(2) 漏洞重现视频/图片，(3) 相关的漏洞修复补丁。需要注意的是，其他漏洞被排除，因为其重现信息不明确（例如，缺少清晰的重现步骤、错误的应用版本或缺少漏洞重现视频/图片）。这一过程耗费了我们大量的人工努力（近六个人月），因为EAIR软件的漏洞重现是 notoriously 很困难。在表2中，#Bugs Selected列给出了这些有效EAIR软件bug的数量。

\subsection{Bug Analysis}
\label{Method}
%\para{Manual investigation for RQ1, RQ2 and RQ3}
%To answer \textbf{RQ1}, \textbf{RQ2} and \textbf{RQ3},
We further manually analyze and label the 885 bugs in terms of underlying causes, symptoms, and modules. Specifically, the underlying causes are determined by manually inspecting the bug descriptions (e.g., error messages) and repair patches. Bug symptoms are manually confirmed by reproducing the corresponding bugs or reviewing the videos or images of reproduction reports. Finally, we find the modules affected by bugs by analyzing system architecture and the directories of bug-related files. 

%为了避免偏差，我们严格约束手工标注过程。
To mitigate potential biases, we strictly regulate the procedure of manual investigation.
Considering that there is a lack of EAIR system bug taxonomy, we leverage the open card sorting approach similar to prior research~\cite{xiong2023empirical} for the taxonomy construction. Specifically, we invite six domain researchers with at least two years of experience and divide them into three pairs. For each pair of researchers, the manual investigation is carried out through an iterative process. In each iteration, each pair of researchers randomly selects 100 bugs and they perform manual analysis individually. According to their domain knowledge, they confirm the underlying cause, bug symptoms, and affected modules for every bug individually. Then, this pair of researchers perform the cross-validation and discuss the bug labels until they obtain a consensus on the classification results. 
We calculated the average Cohen’s Kappa coefficient [15] to measure the consistency of classification results, which is 0.769, thus indicating a decent agreement between the each pair of researchers.
If a consensus cannot be established, we would invite other two pairs of researchers to participate in making the final decision. Noting that such iteration would continue until all of 885 bugs are manually analyzed without omissions. 

\input{tables/symptoms}
\section{RQ1: Symptoms}
\label{EvaluationtoRQ2}

To answer RQ1, we first manually investigate and analyze the surface adverse impact (i.e., specific symptoms) for 885 EAIR system bugs of our dataset.

%在本节中，我们针对数据集中xxx个EAIR软件错误进行分析，以研究其具体表现症状。
%通过分析错误的症状，我们旨在更深入地理解其行为特征及其对软件系统的影响，从而为后续的错误修复提供有价值的见解。同时，不同模块中错误症状的分布情况可以帮助开发者确定维护工作的优先级，使其能够集中精力优化特定模块，提高软件的整体可靠性和稳定性。
%表4列出了所研究项目中错误症状的数量及其占比。

\subsection{EAIR-specific Symptoms }
\label{EAIR-specific cause Symptoms}

We found that 316 (37.62\%) of 840 EAIR system symptoms belong to EAIR-specific Symptoms. We further subdivide them
into 8 subcategories.

\begin{itemize} [leftmargin=10pt]

\item \textbf{S1. Insufficient reliability of AI agent output (RAIO).}
The impact of a bug with such a symptom is that the AI agent produces unexpected outputs, causing the downstream modules not to respond to the agent’s output effectively.

%具有这样的症状的一个bug的影响是AI agent产生难以预料的输出（比如重复的行为），导致下游模块无法响应agent输出。

\item \textbf{S2. Unauthorized access and execution of AI agent (UAEAI).}
A bug with such symptoms can cause the AI agent to unexpectedly initiate operations without obtaining the necessary authorization, or allowing the AI agent to access sensitive resources despite lacking sufficient permissions.
%具有这样的症状的一个bug的影响是 AI agent在没有获得权限的情况下意外地开始执行或 允许AI agent在没有足够权限的情况下访问到敏感资源

\item \textbf{S3. Training failures (TF).}
A bug that exhibits such a symptom may cause the AI agent to fail or crash during the training process, or to learn from incorrectly labeled data and thus produce incorrect outputs.
%具有这样的症状的一个bug的影响是AI agent在训练过程中失败崩溃或AIagent学习到错误的数据产生不正确的输出

\item \textbf{S4. Conflicts and collisions (CC).}
A bug that presents such a symptom may cause EAIRs to encounter conflicts and collisions within the simulation or real-world environments. 
%具有这样的症状的一个bug的影响是机器人在模拟或现实环境中发生碰撞或者冲突

\item \textbf{S5. 3D environment modeling failures (3DEMF).}
A bug with such a symptom can lead to incomplete environment modeling for the EAIR system during simulation, such as object omission in the simulation environment. 
%具有这样的症状的一个bug的影响是机器人软件在仿真环境时出现意外的错误（比如仿真环境遗漏了重要物品，物体悬浮在半空中）

\item \textbf{S6. Poor compatibility of agent (PCA).}
A bug with such a symptom can result in incompatibilities between EAIR and different versions of AI agent models (e.g., GPT versions), or diminish the EAIR agent’s effectiveness when undertaking task planning in large-scale multi-robot scenarios. 
%具有这样的症状的一个bug的影响是EAIR对不同版本AI agent模型不兼容（如GPT版本）, EAIR agent在大规模多机器人场景下进行任务规划效果和兼容性差

\item \textbf{S7. Violation of physical reality (VPR).}
A bug with such symptoms may cause EAIR simulation to produce results that contradict real-world kinematic logic and physical reality. 
%具有这样的症状的一个bug的影响是EAIR产生与真实世界的运动学逻辑以及物理规则相违背的错误（比如物品失重飞行等）

\item \textbf{S8. Embodied control anomaly (ECA).}
A bug that exhibits such symptoms can cause unexpected motions of the EAIR’s components such as the irregular jittering of robot arms. 
%具有这样的症状的一个bug的影响是 EAIRs 系统的组件（如机械臂、摄像头云台、轮子等）出现非预期的运动。

\end{itemize}

In fact, EAIR-specific Symptoms probably reflect the inherent challenges for EAIR system design and development. We take the most common symptoms as examples for illustration. The insufficient reliability of AI agents (RAIO, 32.91\%, 104 of 316) implies the inherent limitation (i.e., hallucination) of AI agents. 
Embodied control anomaly (ECA, 28.16\%, 89 of 316) shows that the control design cannot satisfy the demand on EAIR's behavior learning from the interaction with its surroundings.
Violation of physical reality (VPR, 12.66\%, 40 of 316) probably indicates that EAIR system simulation faces difficulties in accurately modeling real-world dynamics and complex objects. To ensure security and reliability, it is crucial to address the AI agent hallucinations, optimize the EAIR's behavior learning, as well as enhance the sim-to-real logic.

%此外，这些错误症状的出现可能反映出EAIR在设计与实现过程中固有的挑战。例如，the insufficient reliability of AI agent是
%AI agent 的固有局限性，因为AI模型的幻觉。 Motion control anomaly则可能表明开发者在EAIR软件中的控制层设计无法满足AI agent与硬件之间具身交互的需求。
%Violation on physical reality的挑战可能表明EAIR软件仿真模块在建模并遵守真实世界动力学逻辑和复杂物品方面存在困难。为了提升EAIR软件的安全性和可靠性，针对这些领域的深入研究和改进至关重要。这可能涉及AI agent幻觉处理、AI agent的迭代交互方面算法优化、以及仿真到现实的逻辑增强。

%EAIR-specific symptom，这类错误占所有实例的xxx%。其中最常见的错误症状是 Insufficient reliability of AI agent output。
%这反映了EAIR软件在AI agent幻觉处理、AI agent的迭代交互方面、仿真到现实的逻辑设计方面面临挑战。
\begin{tcolorbox}[left=2pt, right=2pt, top=2pt, bottom=2pt]

\textbf{Finding 1:} The EAIR-specific symptoms are common for the bugs in EAIR system(37.62\%, 316/840), which leads to severe functional failures and physical damages. Besides, EAIR-specific symptoms
also reflect the inherent
challenges for EAIR system design and development, such as overcoming the AI agent hallucination, EAIR behavior learning,  designing sim-to-real logic.
\end{tcolorbox}

\subsection{General system Symptom}
\label{Traditional Bug Symptoms in EAIR system}

General system symptoms refer to the regular symptoms similar to the traditional system. 
%We found thatxxx (xxx\%) of xxx EAIR system symptoms belong to general system bugs. 
We further subdivide them
into 7 subcategories.

\begin{itemize} [leftmargin=10pt]

\item \textbf{S9. Crashes.}
When the EAIRs system or its modules are terminated improperly, it may result in a system crash. 
% 当EAIRs系统或组件不当终止时，导致崩溃。

\item \textbf{S10. Hangs.}
A bug that exhibits such symptoms can cause the EAIRs system or its modules to become unresponsive to input while their processes continue running. 
% EAIRs系统或组件无法响应输入，但其进程仍然在运行

\item \textbf{S11. Build errors (Build).}
A bug that exhibits such symptoms can prevent the EAIRs system or its modules from being compiled, built, or installed correctly.
% 阻止EAIRs系统或组件的正确编译、构建或安装。

\item \textbf{S12. Display and GUI errors (GUI).}
A bug exhibiting such symptoms can lead to erroneous outputs, which are displayed in the EAIRs system’s graphical user interface (GUI), visualization components, or human-machine interface (HMI).
%在EAIRs系统的GUI、可视化或人机界面（HMI）中显示错误输出

\item \textbf{S13. Performance degradation (PD).}
A bug exhibiting such symptoms can lead to program performance degradation caused by insufficient code optimization, resource contention, or incorrect algorithm implementation.
%由于代码优化不足、资源竞争、错误的算法实现导致程序运行效率降低

\item \textbf{S14. Compilation failures (Comp.).}
Code syntax errors can cause the program to fail during compilation.
%代码语法错误导致程序编译不通过

\item \textbf{S15. Hardware interaction failures (HIF).}
A bug that exhibits such symptoms can cause failures in the EAIR system’s interaction with hardware
%rendering the hardware unresponsive or even leading to physical damage.
%具有这样的症状的一个bug的影响是EAIR软件与硬件交互时失败,导致硬件无法响应甚至损坏。

\end{itemize}

\begin{tcolorbox}[left=2pt, right=2pt, top=2pt, bottom=2pt]

\textbf{Finding 2:} Among the general system symptoms, the most common ones in the EAIR system are Crash, Hang, and GUI errors, accounting for 18.07\%, 9.35\%, and 14.45\% of all bugs.
\end{tcolorbox}

\section{RQ2: underlying causes}
\label{EvaluationtoRQ1}
After surface adverse impact analysis, we further conduct a underlying cause analysis on 885 EAIR bugs in our dataset.
Specifically, we first establish a taxonomy for EAIR bugs by categorizing each bug according to the EAIR system functionality to which it breaches. Building on this taxonomy, we then carry out a code-level investigation (i.e., manual analysis of bug pattern and repair patch) on these functional bugs to pinpoint the underlying causes. Finally, we reveal the relationship between surface symptom and underlying cause to constitute the foundational steps
toward a comprehensive understanding of EAIR system bugs.

%具体而言，我们首先根据bug对EAIR 功能的违背，针对EAIR bug进行分类以构建分类体系。进一步地，我们在EAIR 功能性bug上进行代码级别的分析（即bug代码和修复补丁），已确定Bug的根本原因。

%Specifically, we successfully identified the underlying cause for the 865 EAIR bugs, 18 EAIR bugs encountered analysis failures due to the lack of debug information and high complexity. 
%With the underlying cause analysis, we establish a taxonomy for EAIR bugs.  

%In this way, we obtain an orthogonal bug taxonomy. Furthermore, the high-level category reflects the functionality violation, and the low-level subcategory refers to the code-level errors.

%在本节中，我们针对数据集中xxx个EAIR bug进行分析，以研究其根本原因。我们成功确定了xxx个功能性错误的根本原因（由于调试信息有限且复杂度较高，我们未能分析其中xxx个错误），并将其归类为三个主要类别：(1) EAIR-specific cause（详见第3.1节），(2) 传统bug（详见第3.2节）。这三个主要类别进一步细分为18个子类别。表3列出了这些根本原因及其分布。

%具体而言，我们遵循操作性错误分类策略（operational bug classification strategy）[32, 60]，以实现正交的分类：归入较高层级子类别的错误不会再被纳入较低层级子类别。此外，较高层级的子类别涵盖更广泛的范围（例如功能级错误），而较低层级的子类别则涉及更具体的范围（例如代码级或变量级错误）。

\input{tables/EAIR_bug_taxonomy}

\input{tables/RootCause_copy}

\subsection{EAIR bug taxonomy}
\label{EAIR-specific-bug}
We summarize the taxonomy for EAIR system bugs, as listed in Table~\ref{taxonomy}.
EAIR system bugs are divided into five categories: (1) category specific to embodied perception, (2) category specific to embodied interaction, (3) category specific to embodied control, (4) category specific to embodied simulation and (5) category others (i.e., the first column). The last category contain traditional robot bugs that are unrelated to EAIR’s four core mechanisms, which have already been analyzed extensively in the prior studies~\cite{} and are therefore not revisited here.
Further, we divide these categories into twelve concrete bug types(i.e., the second column). Each of them reflects the inadequate understanding or incorrect execution of functionality of EAIR's core mechanisms. %For brevity, we aggregate the corresponding underlying causes of these functionality bugs in Table~\ref{taxonomy} (i.e., the fourth column) and provide a detailed exposition in the sections~\ref{rootcauseanalysis}.

%我们总结了如表3所示的EAIR bug 分类法。Category 表示五个主要类别（见第一列），其主要包括具身感知、具身交互、具身控制、具身仿真和其他五个类别。其中，其他类别表示的是与EAIR四个核心机制（即模块）无关的传统机器人bug，这已经在先前的研究中被广泛讨论。由于篇幅限制，我们不再赘述。
%type 表示EAIR bug 的12个具体类型（见第二列），它们反映的是，inadequate comprehension or incorrect execution of EAIR's functionality mechanism。为简洁起见，我们将各类功能性bug的对应的根本原因集中列于表3中，并在下文对其进行进一步解释。

As can be seen, 657 of 885 EAIR system bugs (74.24\%) are the specific bugs related to EAIR's four core mechanisms (i.e., embodied perception, embodied interaction, embodied control, embodied simulation). Within this subset, bugs stemming from the embodied-simulation category are the most prevalent, constituting 230 of the 657 bugs (35.01\%).
These statistics indicate that developers struggle to comprehend and implement the EAIR's core mechanisms correctly, with the inherent complexity of the embodied simulation posing the greatest challenge.
Focusing on the first four categories, physical-feature modelling incompleteness and machine-learning issues emerge as the most prevalent functionality bug types, accounting for 168/885 (18.98 \%) and 139/885 (15.71 \%) of the bugs, respectively. These findings show that modelling sophisticated physical features like fluid dynamics and softbody physics and introducing embodied agents create significant reliability issue in EAIR systems.

%可以看到，885个EAIR system bugs中的664个 (xx%) 是与EAIR四个核心机制（即模块）相关的bug, 其中具身仿真相关的Bug占比是最多的（即xx%，241/664)。这表明开发人员很难正确地理解和实现EAIR系统的四个核心机制，造成了大量的系统bug，而复杂的具身仿真模块的给EAIR系统的实现带来了最大的挑战。而在前四个类中，Physical-feature modeling incompleteness 和 Machine learning issue 是占比最多的bug 类型，分别占20.90% (185/885) 和 16.61% （147/885）。这表明，针对像fluid dynamics and softbody physics 的高级物理特征的建模和embodied agent 的引入，容易给EAIR系统带来了严重的安全问题

\begin{tcolorbox}[left=2pt, right=2pt, top=0pt, bottom=0pt]
%\small
\textbf{Finding 3:} Among the 885 EAIR bugs, 657 (74.24\%) of them are attributed to errors specific to EAIR core mechanisms (i.e., embodied interaction, embodied control, embodied simulation and embodied precision), which are primarily caused by inadequate comprehension or execution of these mechanisms.
\end{tcolorbox}

\subsection{Underlying cause analysis}
\label{rootcauseanalysis}

Further, for each type of functionality bugs in the second column of Table~\ref{taxonomy}, we conduct manual analysis on their bug patterns and repair patches for finding out the underlying cause (i.e., code-level error). 
And Table~\ref{crosstable_transposed_updated} lists all of the underlying causes and
their statistical distribution.
As can be seen, the underlying causes of EAIR bugs can be divided into two parts including EAIR-specific causes and traditional causes.

\para{EAIR-specific causes} 
We found that 118 (13.33\%) of 885 EAIR system bugs belong to EAIR-specific causes.
While EAIR-specific causes constitute a relatively small proportion, they remain extremely significant, as they can lead to severe damage, as shown in Section~\ref{Cause and Symptoms}.

%与EAIR-specific cause是指与EAIR特定功能机制相关的错误，这类错误通常由对EAIR独特机制理解不足和/或不正确的执行方式引发。我们发现，在xxx个EAIR软件bug中，有xxx个（占比xxx%）属于与EAIR特定的错误。虽然 EAIR-specific causes 占比不多，但是他们仍然是极为重要的，他们会产生严重的功能性不良的症状，产生较大的危害，正如我们在5.3节展示的那样。
%根据涉及的EAIR独特机制，我们进一步将其归类为四个主类别互斥的8个子类别。

%AI agent输出的完整性（即它们是否被攻击者更改），这可以直接影响机器人行为的正确性,安全违规（如碰撞），规则违反，移动性下降（如识别障碍错误)

\begin{itemize} [leftmargin=10pt]
    \item \textbf{C1. Integrity of AI agents output (IAIO).} This cause can directly impact the behavior correctness of a given robot, such as security violation (e.g., collisions), adversary manipulation, and mobility reduction (e.g., obstacle recognition error).
    %As shown in Figure 3(a), when the AI agent conducts the initialization process (lines 1–3), it ignores the robot’s body-part information, as addressed by the patch in lines 6–11. In this case, the agent was unable to obtain valid input and consequently failed.
    %如图3（a）所示，由于AI agent在初始化机器人状态时（line1-3）忽略了机器人各部位信息如补丁（line 6-11）所示，导致AI agent无法获得有效输入而失败

    \item \textbf{C2. Permission leaks of AI agents (PLAI).} This cause refers to the leakage of the agent’s planning information, privacy-sensitive location data, and critical permissions.
    %As illustrated in Figure 3(b), since the program omits the access control constraint on write permissions (as added in the patch of lines 3–5), the AI agent can access a critical system resource (line 6) without sufficient permissions, causing a bug. 
    %如图3（b）所示，AI agent 在访问系统的关键资源（line 6），缺少必要的写入权限，而程序中缺少对写入权限的访问控制检查导致bug，如补丁所示（line 3-5）

\item \textbf{C3. Availability of AI agents (AAI).}
A typical feature of EAIR is its compound AI agents, where downstream AI agents depend on the timely and reliable outputs of upstream AI agents. Poor availability (i.e., the ability for reliable outputs) can result in denial of service or outcomes that violate physical reality.

%AI模型的错误和训练数据中的错误标签导致agent运行崩溃或训练失败
\item \textbf{C4. AI agent model and dataset issue (AIMD).}
Errors in the AI model can cause the agent to crash. For example, the misalignment between the AI model's output and the downstream component’s input can cause a crash. 
%As shown in Figure 3(c), the AI agent crashes during forward pretraining due to an incorrect masking rule (line 3). This bug can be resolved by correcting the masking rule (lines 4–5).
%如图3（c）所示，AI agent在前向预训练时，由于掩码规则错误（line 3）导致训练崩溃，这种bug可通过纠正掩码规则来解决（line 4-5）

%EAIR的另一个典型特征是多模态信息处理，它要求对多种模态信息（如图像，声音等）整合，以帮组agent进行高级任务规划。这种整合涉及复杂的程序逻辑，容易导致错误。
\item \textbf{C5. Multi-modal information aggregation error (MIAE).}
Another typical feature of EAIR lies in its capacity for multi-modal information processing(e.g., images, audio). This integration process involves complex program logic, making it prone to errors.
%As presented in Figure 3(d), the AI agent makes an error during training by integrating only the environmental data (line 4) and ignoring the robot’s information (line 6 in the patch). This omission leads the AI agent to learn incorrect data.
%如图3（d）所示，AI agent在训练时，整合多模态信息时出现错误，只合并了环境信息（line 4），而忽略机器人本身信息（patch 的line 6），导致AI agent学习错误数据。

%\item \textbf{C6. Multi-modal information sensing error.}  

%区别于传统机器人软件（如自动驾驶），EAIR的仿真需要对更加复杂的交互环境（如3D复杂加剧环境）和运动学逻辑（如物品冲突）进行建模。仿真模块由于针对物理世界模拟的逻辑不完善导致的bug，我们称之为 Sim-to-real rule inconsistency。
\item \textbf{C6. Sim-to-real rule inconsistency (SRI).}
%Unlike traditional robotic systems (e.g., autonomous driving systems), EAIR simulation requires more intricate interactive environments (e.g., complex 3D domestic scenarios) and kinematic logic (e.g., object collisions (e.g., object collisions). 
We refer to bugs arising from inadequate or flawed physical-world modeling within the simulation module as Sim-to-real rule inconsistency.
%As illustrated in Figure 3(f), an incorrect rule vector (line 4) is used in the simulation module, causing the light rays to be emitted in the opposite direction from what would occur in a real-world setting. This discrepancy leads to an inconsistency between the simulation and practical conditions.
%如图3（f）所示，在仿真模块中，使用了错误的规则向量（line 4），导致有向光以真实场景相反的方向射出，产生仿真与现实的不一致。

%区别于传统机器人，EAIR拥有更复杂的架构和模块。这种bug是指仿真模块中由于EAIR架构和模块建模不完整或不合理导致的问题
\item \textbf{C7. Embodied simulation model error (ESME).} 
%In contrast to traditional robots, EAIR is characterized by a more intricate architecture and modular design. 
This bug refers to issues within the simulation module that arise from incomplete modeling of the EAIR architecture and its modules.
%As shown in Figure 3(e), in the EAIR's simulation module, the simulation model (i.e., the URDF file) contains incorrect parameters, preventing the model from behaving as intended within the simulated environment.

% 如图3（e）所示，EAIR在仿真模块中，仿真模型（即urdf文件）出现参数错误，导致仿真模型无法在仿真环境中实现预期行为。

%这个问题指的是AI agent层，控制层，桥接层和硬件之间的交互过程中的错误
\item \textbf{C8. Embodied module interaction issue (EMII).} This issue refers to the interaction-related bug among the embodied agent, core-mechanism layer, bridge layer and hardwares.

\end{itemize}

%在393个EAIR bugs中，有xxx%是由与EAIR机制（比如AI agent，具身交互，多模态信息整合和 sim-to-real adaptation）相关的错误引起的，这主要是由于对EAIR机制的理解不够充分所导致。其中，AI agent的占比是最高的，子类别也是最多。

\begin{tcolorbox}[left=2pt, right=2pt, top=0pt, bottom=0pt]
%\small
\textbf{Finding 4:} Among the 118 EAIR-specific causes, AI agent's bugs possess the most kinds of causes (i.e., 4) and largest proportion (i.e., 56.77\%, 67/118).
\end{tcolorbox}

%\subsection{General Programming Bug}
%\label{General Programming Bug}

%General programming bugs are caused by the typical errors that are encountered in the classic system development process. 
%We found that xxx (xxx\%) of xxx EAIR system bugs belong to general programming bugs. 
%We further subdivide the general programming bugs into 10 subcategories.

%一般编程错误（General Programming Error）指的是在经典软件开发过程中经常出现的错误。我们发现，在393个功能性错误（functional bugs）中，有223个（占比56.7%）由一般编程错误引起。我们进一步将其归类为七个不同的根本原因子类别（详见表3）。在下文中，我们将逐一介绍表3所列出的各个子类别。

\input{tables/causeandsymptom}

\para{Traditional causes} Traditional causes are the typical errors that are encountered in the classic system development process.

\begin{itemize} [leftmargin=10pt]
%此根本原因涉及到不正确的数值计算、值或使用。
\item \textbf{C9. Incorrect numerical computation (INC).} This bug is caused by incorrect computation logic between variables.

%一个或多个变量被错误地赋值或初始化。
\item \textbf{C10. Incorrect assignment (IA).} This cause refers to unexpected initialization or false assignment of variables. 

%不正确赋值也是EAIR软件错误的常见根本原因之一，在xxx个错误中占比xxx%。
%\begin{tcolorbox}[left=2pt, right=2pt, top=2pt, bottom=2pt]
%\textbf{Finding 3:} Incorrect assignment is the most common underlying cause for EAIR system bugs, accounting for 174 (19.71\%) of 885 EAIR system bugs.
%\end{tcolorbox}

%缺少必要的条件判断语句。
\item \textbf{C11. Access control incompleteness (ACI).} This bug arises from the lack or misuse of access control checks.

%数据结构定义错误、数据结构指针被误用，或类型转换不正确
\item \textbf{C12. Incorrect data structure (IDS).} This cause refers to errors in data structure definitions, misuse of data structure pointers, or incorrect type conversions.

%此根本原因涉及到错误使用其他系统或库的接口（例如，已弃用的方法、参数设置错误等）。
\item \textbf{C13. Misuse of an application programming interface (MAPI).} This cause includes misuse of external APIs from other systems or libraries (e.g., invoking an outdated library) or misuse of internal APIs from other components (e.g., incorrect call chains).

% 此根本原因涉及到误用其他组件的接口——例如，调用顺序不匹配；违反继承契约；以及错误的打开、读取和写入操作
%\item \textbf{C15. Misuse of an internal interface}

% 由于条件表达式错误导致的漏洞
%\item \textbf{C16. Incorrect condition logic}

%此根本原因涉及错误使用并发相关的结构（例如，锁、临界区、线程等）
\item \textbf{C14.Concurrency issue (Con.).} This bug is caused by misuse of concurrency-related structures such as locks and threads.

%此根本原因涉及内存的误用（例如，不正确的内存分配或释放）。
\item \textbf{C15. Memory issue (MEM.)} This bug arises from unexpected usage of memory such as erroneous memory allocation or release.

%此根本原因涉及手册、教程、代码注释和未被AV系统执行的文本不正确。
\item \textbf{C16. Invalid Documentation (ID).}
This underlying cause involves errors present in manuals, tutorials and code comments.

%
%\item \textbf{C17. Incorrect configuration.} This bug is caused by erroneous modifications in files related to compilation, building, compatibility, and installation.

%\para{C21. Logical deficiency}

\item \textbf{C17. Cyber Configuration Error (CCE)} This underlying cause reflects the cyber failure of the EAIR system

\item \textbf{C18. Other Issue} These bugs refer to import issues, spelling errors, unimplemented functionality and logging errors.

\end{itemize}

%\begin{tcolorbox}[left=2pt, right=2pt, top=0pt, bottom=0pt]
%\small
%\textbf{Finding 1:} Among the 118 EAIR-specific causes, AI agent's bugs possess the most subcategories (i.e., 4) and largest proportion (i.e., 56.77\%).
%\end{tcolorbox}

%EAIR软件错误具有多样化的根本原因，这些原因可归类为五个主要类别，并进一步细分为15个子类别
\begin{tcolorbox}[left=2pt, right=2pt, top=2pt, bottom=2pt]
\textbf{Finding 5:} The underlying causes of EAIR system bugs can be divided into 18 kinds. Incorrect assignment (IA) is the most common underlying cause, accounting for 174 (19.66\%) of 885 EAIR system bugs.
\end{tcolorbox}

\subsection{Relationship between Cause and Symptom}
\label{Cause and Symptoms}

For developers and researchers in the domain of EAIR, a thorough understanding of the relationships among a given underlying cause, a specific bug symptom, and the frequency with which the given underlying cause leads to the specific bug symptom is essential for guiding the avoidance, detection, and repair of EAIR system bugs.

%对于从事EAIR研发与工程工作的技术人员与研究者而言，深刻理解某一特定根本原因导致特定症状的频率三者之间的相互关系，对于预防、检测、定位并修复EAIRbug具有重要的指导意义。基于这一考虑，我们进一步考察了
%What kinds of bug symptoms can each underlying cause produce?

%表6展示了在EAIR软件中，特定根本原因导致特定症状的程度。值得注意的是，在我们的分类方案中，不正确的赋值是最常见的根本原因。毫不意外地，该原因引发了多种症状，覆盖了我们分类方案中的15种症状中的13种。
%其中，不正确赋值实现尤为显著地影响了Display and GUI errors（出现41次）和 Insufficient reliability of AI agent output（出现28次），Crash（出现17次）和 Conflicts and collisions（出现16次）。 这表明相较于自动驾驶的其他方面，变量的定义与赋值的容易带来更广泛的影响。除此之外，由于不正确的赋值实还频繁引起 Insufficient reliability of AI agent output（出现28次）和Conflicts and collisions（出现16次）这些EAIR特定的症状，这表明EAIR特定的症状并不一定由EAIR特定的underlying cause（即EAIR特有机制的错误）导致，还会由EAIR特有机制于其他部分软件组件交互过程中General Programming Bug导致

Table~\ref{causeandsymptom} presents the frequency of symptoms that each underlying cause may exhibit in the EAIR system. Notably, incorrect assignment (IA) emerges as the most prevalently occurring underlying cause, which covers 13 out of the 15 symptoms included in our symptom classification. %Specifically, the most common symptoms caused by the incorrect assignment are “Display and GUI errors” (41 instances), “Insufficient reliability of AI agent output” (28 instances), “Crash” (17 instances), and “Conflicts and collisions” (16 instances). 
This observation shows that variable definition and assignment can have a far-reaching influence compared to other underlying causes of EAIR bugs.
Note that incorrect assignment (IA) frequently leads to EAIR-specific symptoms such as “Insufficient reliability of AI agent output” (RAIO, 28 instances) and “Conflicts and collisions” (CC, 16 instances). This indicates that, except for EAIR-specific underlying causes (i.e., EAIR-specific mechanism errors), EAIR-specific symptoms may also stem from traditional causes, which occur during the interaction between EAIR-specific mechanisms and other components.

%\begin{tcolorbox}[left=2pt, right=2pt, top=2pt, bottom=2pt]
%\textbf{Finding 7:} Incorrect assignment is the most frequently occurring underlying cause, which
%cause resulted in a wide variety of symptoms, producing 13 out of
%15 symptoms in our classification scheme.
%\end{tcolorbox}

As shown in Table~\ref{causeandsymptom} (rows 2–9), EAIR-specific underlying causes primarily give rise to EAIR-specific symptoms. For instance, both "Availability of AI agents (AAI)" and "AI agent model and dataset issue (AIMD)" belong to AI agent bugs, and all mainly lead to AI-agent-related symptoms such as the "insufficient reliability of AI agent output (RAIO)" (60\%, 6/10 and 53.33\%, 16/30 of cases, respectively). 
%Meanwhile, Sim-to-real rule inconsistency and Embodied simulation model error both belong to Sim-to-real adaptation bugs and mostly cause simulation-related symptoms such as violation of physical reality (41.67\%, 5/12 and 80\%, 4/5 of cases, respectively).

\input{tables/compositeBug}

In addition, we observe that EAIR-specific underlying causes mainly lead to EAIR-specific symptoms that affect EAIR’s motion (e.g., "conflicts and collisions (CC)", "3D environment modeling failures (3DEMF)", "violation of physical reality (VPR)", and "embodied control anomaly (ECA)"). Such symptoms may result in unexpected behaviors of EAIR within real-world or simulation environments, leading to anomaly behavior  or even physical damage.

%我们可以看到， EAIR specific underlying cause （表3 2-9行）主要会导致与EAIR特定机制相关的漏洞。比如， Availability of AI agents 和 AI agent model and dataset issue都属于AI agent bug， 都主要主要导致的症状是 Insufficient reliability of AI agent output （分别占60%和50%）。Sim-to-real rule inconsistency 和 Embodied simulation model error 都属于Sim-to-real adaption bug， 都主要导致 Violation on physical reality。 
%进一步地，我们发现，EAIR specific underlying cause直接导致了许多影响EAIR行动的症状（例如 Conflicts and collisions 、 3D environment modeling failures 、 Violation on physical reality 、and  Embodied control anomaly）。这些症状可能造成EAIR在现实或仿真环境中表现unexpected的行为，造成功能性的失败（如xxx）或物理损失。

\begin{tcolorbox}[left=2pt, right=2pt, top=2pt, bottom=2pt]
\textbf{Finding 6:} EAIR-specific causes mainly contribute to 4 types of EAIR-specific symptoms, which reveals a correlation relationship between EAIR-specific symptoms and EAIR-specific underlying causes.
\end{tcolorbox}

%进一步地，我们调查了bug与symptom的对应关系，并将它们的对应关系划分为三种类型：T1. Bugs that one symptom are caused by multiple causes; bugs that one cause can lead to multiple symptoms 和 bugs that cause与symptom一一对应。注意，其中T1和T2代表复合bug类型。表4展示了我们的进一步调查的结果。可以看到，EAIR-specific cause 中复合bugs占比远远高于 General Programming Bug。这表明EAIR-specific cause拥有更高的复杂度，更难以被发现。同时也暗示了，可能有很多的EAIR-specific cause尚未被发现。

Furthermore, we investigated the correspondence between bugs and symptoms and categorized their relationships into three types: T1, T2, and T3. Specifically, T1 represents scenarios in which a single symptom is caused by multiple underlying causes, T2 refers to cases where a single underlying cause leads to multiple symptoms, and T3 indicates a one-to-one mapping between cause and symptom. Notably, T1 and T2 constitute compound bug types. Table 4 presents the results of our additional investigation. It can be observed that, among EAIR-specific causes, the proportion of compound bugs is significantly higher than that among Non-compound bugs. This finding indicates that EAIR-specific causes exhibit greater complexity and are difficult to detect.
%and it further implies that numerous EAIR-specific causes may currently remain undiscovered.

\begin{tcolorbox}[left=2pt, right=2pt, top=2pt, bottom=2pt]
\textbf{Finding 7:} 79 (66.95\%) of 118 EAIR-specific causes are compound bugs. Among them, 47 belong to bugs whose one symptom is caused by multiple causes, and 32 belong to bugs whose one cause can lead to multiple symptoms.
\end{tcolorbox}

\input{tables/causeandmodulenew}

%\begin{figure}[t]
%\centering
%\includegraphics[width=3.4in]{figures/EAIR_architecture.pdf}
%\caption{A layered map of bugs in our summarized EAIR system architecture.} 
%\label{architecture}
%\vspace{-2mm}
%\end{figure}

\section{RQ3: Module distribution}
\label{EvaluationtoRQ3}

In this section, we investigate the distribution of bugs across different EAIR modules. 
Specifically, we first investigate the relationship between underlying cause and modules, to indetify  which modules are inherently
more bug-prone or play a pivotal role in bug identification
and repair. For researchers, these information can pinpoint
the system parts that merit further efforts in bug prediction,
detection and repair.
Then, we investigate the relationship between symptom and modules, to facilitate engineers to allocate
more resources(e.g., development time and effort) toward
modules which possess the most severe errors or the greatest
latent risk. Furthermore, according to Conway’s Law [31], this
relationship can also inform managers and technical leaders as
to how different bug symptoms will affect different teams of
an EAIR system development.

\subsection{Relationship between Underlying Cause and EAIRs Modules}
\label{Bug Cause Occurrences in EAIRs Modules}
In this subsection, we aim to conduct a thorough
understanding of the relationships among a given underlying cause, a
specific module, and the frequency at which the given root
cause affects the specific module, and the analysis results are presented in Table~\ref{causeandmodule}.

%module only 
%我们发现每个module都包含多种bug（即underlying cause）而且差异较大，这给软件开发和维护带来很大挑战。其中， Simulation is the most frequent module that occurs bug, which occurs 17 out of 18 underlying causes in our classification。这表明（1）仿真模块由于其需要尽可能地模拟现实世界的复杂物品和逻辑，固然较其他模块在实现方面的复杂度更高；（2）当前仍然缺乏足够可靠且有效的仿真模块或软件，其开发仍然是任重道远。
\para{Analysis on module perspective} When considering the modules individually, We found that each module comprises multiple bugs (i.e., underlying causes) with significant variation among them, which presents critical challenges for bug detection and repair. Notably, the simulation module is the most frequent module that occurs bug, which occurs in 17 of 18 underlying causes in our classification. It is fairly straightforward to understand. The simulation is inherently more complex in its implementation, given its demand to accurately simulate the complex objects and logic of the real world.

%\begin{tcolorbox}[left=2pt, right=2pt, top=2pt, bottom=2pt]
%\textbf{Finding 9:}Simulation is the most frequent module that occurs underlying cause, which occurs 17 out of 18 underlying causes in our classification
%scheme. 
%\end{tcolorbox}

%underlying cause only
\para{Analysis on underlying cause perspective}
%When considering the underlying causes individually, We found that the module distribution of EAIR-specific causes is significantly different from that of traditional causes, hence we discuss them separately. 

Among the EAIR-specific causes, we observe two key points: (1) each underlying cause tends to appear in only a limited number of modules (typically 2 to 6), and (2) the affected modules are primarily those with functionality dependencies on the EAIR-specific mechanisms being violated.
For instance, the underlying cause of \textit{ AI agent model and dataset issue (AIMD)} predominantly arises in the Learning module, where it appears 20 times, accounting for 64.52\% of all instances. The Simulation, Reasoning, and Grasping modules also exhibit this underlying cause, with 6, 3, and 2 instances respectively. Since the AI agent model and dataset issue violates the implementation of AI agents, it directly affects modules that employ the AI agent (i.e., Learning and Reasoning) and indirectly impacts modules that depend on the AI agent’s output (i.e., Grasping and Simulation).

%%我们发现EAIR-specific causes与General system bugs差异较大，因此我们分开讨论。Among EAIR-specific cause, when considering the symptoms individually, 我们可以发现：一是每种underlying cause （即bug type）主要仅出现在少数几个模块中（通常为 2 到 5 个模块）；二是这些underlying cause通常对EAIR-specific机制的违反，因此underlying cause主要影响与被违反机制在功能上存在依赖的模块。
%举例， AI agent model and dataset issue的underlying cause主要出现在 learning 模块中，该模块出现了 20 次，占该症状所有实例的 xxx%。其次是 simulation 模块， Decision 模块和 motion模块，分别出现 6 次，3 次和 2 次。其中， AI agent model and dataset issue违反了EAIR中AI agent的实现，这直接地影响了使用了AI agent模块 （即learning 和decision），同时也间接影响依赖于AI agent输出的模块（即motion和simulation）

%Planning 模块负责生成用于EAIR仿真的虚拟环境，而Motion 模块和 Decision 模块则与仿真环境中的机器人行为和移动密切相关。这些模块协同工作以确保EAIR的准确仿真，任何一个模块中的不准确都可能引起 EAIR仿真到现实不一致。

\begin{tcolorbox}[left=2pt, right=2pt, top=2pt, bottom=2pt]
\textbf{Finding 8:}
Among EAIR-specific causes, each of them has a different
set of dominant modules. And each underlying cause distribution is always closely related to the modules’ functionalities.
\end{tcolorbox}

%EAIR-specific cause

%General programing bug
%

Among traditional causes, when considering the underlying causes individually, we can observe modules affected by each underlying cause vary. Misuse of API (MAPI) and Access control incompleteness (ACI) are the most frequent underlying causes that affects modules, which affect all 13 modules in our EAIR architecture.
EAIR systems are designed to facilitate extensive invocations among AI agents, other system modules, and hardware devices, which necessitate the usage of complex internal and external APIs and diverse access control implementation. Consequently, such systems are particularly susceptible to API misuse and access control incompleteness. This observation implies that invocation bugs in EAIR systems (e.g., API misuse) represent an important research direction that merits deeper exploration in future studies.
%这实际上和EAIR-specific特有的机制（即具身交互）有关，EAIR设计AIagent与其他软件组件，软件与硬件的广泛交互，因此使用了大量的内部与外部API，因此容易发生API误用。这暗示了EAIR的交互bug(如API误用)是一个未来值得深入探索的问题。

\begin{tcolorbox}[left=2pt, right=2pt, top=2pt, bottom=2pt]
\textbf{Finding 9:} Invocation bugs in EAIR systems
represent an important research direction
that merits deeper exploration in future studies.
\end{tcolorbox}

%跨模块bug指的是存在于多个模块之间交互的bug。进一步地，我们对跨模块bug的underlying cause进行了分析。从表xx可以看出，xxx、xxx、xxx是最常见的跨模块bug。这些缺陷通常源于EAIR软件中各模块之间的相互依赖性。
%Cross-module bugs refer to specific bugs that occur during interactions between multiple modules. Furthermore, we analyzed the underlying causes of these cross-module bugs. As shown in Table~\ref{cross-module bugs}, misuse of API, incorrect assignment, and sim-to-real rule inconsistency are the most common cross-module bugs. These bugs typically arise from the dependencies between the multiple modules within the EAIR system.
%For example, the sim-to-real rule inconsistency typically involves the motion, simulation, and navigation modules, reflecting explicit inter-module dependencies. In particular, the simulation module is heavily dependent on movement data (e.g., global variables) provided by the motion and navigation modules. Consequently, a bug that affects such shared global variables can propagate across modules, resulting in cross-module bugs. 
% multi-modal information aggregation error 通常涉及 sensing，interaction和 3D map construction等模块，而这些模块存在明显的依赖关系，即3D map construction模块明显依赖于sensing，interaction感知到的环境信息。一旦存在一个bug能影响到这些全局变量，便会引起跨模块bug

%总结而言，由于自动驾驶软件的复杂性和模块化特性，通常需要跨多个模块进行修改和调整，才能有效检测修复缺陷。

%\begin{tcolorbox}[left=2pt, right=2pt, top=2pt, bottom=2pt]
%\textbf{Finding 10:}
%The most common underlying causes of cross-module bugs are API misuse (18.97\%, 11/58), incorrect assignment (17.24\%, 10/58), and sim-to-real rule inconsistency (8.62\%, 5/58). 
%\end{tcolorbox}

%\input{tables/cross-module}

\subsection{Bug Symptoms in EAIRs Modules}
\label{Bug Symptoms in EAIRs Modules}

Furthermore, we also conduct a thorough analysis on
the relationships among a given symptom, a specific module, and
the frequency which the given symptom exhibits in the specific module and the analysis results are shown in Table~\ref{symptomandmodule}.

\para{Analysis on module perspective}
%
%我们发现每个module都包含多种bug 症状而且不同模块之间存在较大的差异，这给软件开发和维护带来很大挑战。其中， Simulation， configurator, and learning are the most frequent module that exhibits bug symptoms, where the Simulation module produces 14 out of 15 bug symptoms as well as configurator and learning modules exhibit 12 out of 15 bug symptoms in our classification。这表明（1）仿真模块和learning模块容易受到广泛的bug不良症状的影响，开发和维护人员应该用更多和更仔细的努力来提升这些模块的可靠性和安全性（2）当前EAIR的learning和simulation依然是最不可靠和最不安全的组件。
%
When considering the modules
individually, we observed that each module encompasses multiple bug symptoms, and significant differences exist among various modules, which poses considerable challenges to EAIR system development and maintenance. Among these modules, Simulation, Configurator, and Learning are the most frequent modules that exhibit bug symptoms. For example, the simulation module produces 14 out of all 15 bug symptoms, configurator and learning modules exhibit 12 out of all 15 bug symptoms. These results imply that the simulation, configurator, and learning modules are especially prone to adverse symptoms and thus require more meticulous efforts and future research to improve their reliability and safety.

\para{Analysis on symptom perspective}
%We also found that the module distribution of
%EAIR-specific symptoms are significantly different from those of general system symptoms, hence we discuss them separately
%
Among EAIR-specific symptoms, when considering each symptom
individually, we observe two key points: (1) embodied control anomaly (ECA) emerges as the most prevalently occurring symptom, which covers 12 out of the 13 modules included in our EAIR modules classification. (2) a major part of the distribution for a given symptom is always closely related to the functionalities of the corresponding modules. For example, most of the symptoms (43 out of 83) for embodied control anomaly (ECA) occur in the Grasping module. 
These observations indicate that EAIR-specific symptoms exhibit a high degree of locality within the modular structure of EAIR systems. In other words, the occurrence of a particular symptom is closely tied to the responsibilities and operations of specific modules. This insight can guide efforts aimed at detecting or fixing specific types of bugs. For instance, one could leverage pruning-based approaches to optimize the efficiency of bug detection by concentrating on the most relevant modules.

\begin{tcolorbox}[left=2pt, right=2pt, top=2pt, bottom=2pt]
\textbf{Finding 10:} Similar to the underlying cause,  each symptom distribution is also always closely related to the modules’ functionalities, which can guide efforts aimed at detecting or fixing specific types
of bugs, e.g., efficient pruning-based bug detection.

\end{tcolorbox}

\input{tables/symptomandmodulenew}

\section{IMPLICATIONS AND DISCUSSIONS}
\label{implications}
%在本节中，我们讨论了本研究（RQ1～RQ4）所得出的启示，并阐明了开发者和研究人员应如何应对 Android 应用中的功能性缺陷。

\subsection{Implications}
\label{implication}

\para{Mitigating EAIR Bugs in Development}
With the symptom analysis (RQ1), we observed that (1) EAIR-specific symptoms are common in EAIR system bugs and can lead to functional failures and severe damage (Findings 1 and 2). Hence, developers should pay considerable attention to high-impact EAIR-specific symptoms and leverage the correlation information we summarized for the eight EAIR-specific symptoms to facilitate EAIR bug diagnosis (see Section~\ref{EAIR-specific cause Symptoms}).

%finding2 and finding1 (避免EAIR-specific机制的错误)
%finding3 （见下）
%不正确赋值是影响众多EAIRbug的最常见根本原因之一，占比达 19.71%（发现 3，RQ1）。这表明，开发者在开发或测试特定模块功能时，还应足够谨慎地初始化或修改关键变量的赋值。例如，对于EAIR的motion和Navigation模块，开发者应仔细考虑与机器位置相关的重要变量的初始化和修改。事实上，我们对本研究中EAIR systems的开发者文档进行检查后发现，其中有许多开发者文档（xx%）在变量定义和初始化赋值上并不足够准确。
%进一步地，对于EAIR-specific causes，我们发现，（1）他们主要是对 AI agents, embodied interaction, multi-modal information aggregation, and sim-to-real adaptation四种机制的违反 （发现 2，RQ1）；（2）AI agents bug 在它们中间有着最大的占比 （发现 1，RQ1）。因此开发者可以谨慎地理解这四种EAIR机制（尤其是第一种）并确保程序中的关键语义的正确实现以避免此类不良编程实践。
%
Within the underlying cause analysis (RQ1),
incorrect assignments rank among the most common underlying causes affecting numerous EAIR bugs (Finding 5). This observation indicates that developers should exercise more caution when initializing or modifying critical variables during the development or testing of module functionalities. For instance, in the Grasping and Navigation modules of EAIR systems, particular care must be taken regarding the initialization and modification of key variables related to robot position. Indeed, by investigating the developer documentation for the EAIR systems, we discovered that many EAIR systems (53.75\%, 43/80) lack sufficient accuracy in defining and initializing the key variables.
Furthermore, in terms of EAIR-specific causes, we found that (1) they primarily stem from violations of four EAIR-specific mechanisms, i.e., embodied perception, embodied interaction, embodied control, and embodied simulation (Finding 3); and (2) AI agent bugs constitute the largest proportion among these (Finding 4). Consequently, developers should ensure a thorough understanding of these four EAIR mechanisms (particularly AI agents) and maintain correct implementation of their key semantic elements to avoid  incorrect programming practices.

%finding4 和 finding5 （见下）
%finding 6 （见下）
% 通过分析缺陷症状（RQ2），我们发现：（1）The EAIR-specific symptoms are common for the bugs in the EAIR system(42.31%, 344/813）， which can lead to functional failures and severe damages （发现 5和6， RQ2）。因此，开发者可以重点关注危害大的EAIR-specific symptoms，并利用我们总结8种EAIR-specific symptoms相关性信息来辅助诊断EAIR bug (Section 5.1)。

%finding 8 （有针对性避免module错误）
%The most bug-prone modules are simulation (27.29%), learning (16.65%), and motion (12.46%) （发现 8， RQ3））。这表明，这些模块require more focused and meticulous efforts to improve their reliability and safety.事实上，我们通过对开发者文档中这些模块的描述进行检查后发现，其中有大部分EAIR系统（xx%）在开发的时候在这些模块上存在简单重用传统软件的第三方库或包，容易忽略实现EAIR-specific mechanisms。

And module distribution analysis (RQ3) pinpoint the most bug-prone modules, which underscores more meticulous efforts.

%\label{BugDetecting}
%（利用pattern进行漏洞检测）
%本研究显示（表xx），我们为18种EAIRbug总结了xxx中bug-patch patterns。特别的是EAIR-specific causes form more concentrated clusters, and the EAIR-specific causes of each cluster exhibit relatively high similarity. 因此，我们可以选择合适的pattern加以总结和推理去发现EAIR系统中的bug。正如我们在7.2节展示的那样，我们证实了结合了相应的pattern的静态分析一发现EAIR bug的可行性。据我们所知，这是目前首个用于分析EAIR bug的工具。
%
%As mentioned in Table~\ref{}, we have summarized xxx bug-patch patterns for 18 types of EAIR bugs. Notably, EAIR-specific causes form more concentrated clusters, and each cluster’s EAIR-specific causes exhibit relatively high similarity. Consequently, researchers can adapt appropriate patterns and utilize them to infer and summarize bug detection in EAIR systems. As demonstrated in Section~\ref{}, we establish a POC tool to identify EAIR bugs via static analysis that is enhanced with these patterns, and experiment results confirm the effectiveness of these patterns. To the best of our knowledge, our POC tool is the first of its kind for EAIR bug analysis.

%finding7 and finding 10 （复杂漏洞检测）
%本研究显示， EAIR存在复杂bug：（1）79 (68.10%) of 118 EAIR-specific causes are compound bugs (finding7, RQ2). （2）EAIR存在相当数量(58)的跨模块bug (finding10, RQ3).
%Actually, such types of bugs exhibit greater complexity and are more difficult to discover.  This observation implies an important research direction that merits deeper exploration in future studies.
\para{Detecting and Repairing EAIR Bugs after Deployment}
Our study indicates that EAIR systems exhibit complex bugs in two major aspects: (1) among the 118 EAIR-specific causes, 79 (66.95\%) are compound bugs (Finding 7 in RQ2); and (2) there is a considerable number (i.e., 58) of invocation bugs occurring in EAIR system (Finding 9 in RQ3). In practice, such bugs demonstrate greater complexity and are more difficult to detect. This observation suggests an important research direction that warrants in-depth exploration in future studies.

%finding9和finding11 （分析效率优化）
%此外，通过对underlying cause和 symptom 的模块化分析，我们发现，对于 Among EAIR-specific causes, each of underlying cause and symptom has a different set of dominant modules. And their distribution is always closely related to the modules’ functionalities. 借助这些发现，我们可以通过聚焦在最相关的模块来进行bug的检测和修复，从而实现分析效率的提升和优化。

Furthermore, with a modular analysis of underlying causes and symptoms, we found that, among EAIR-specific causes, each of the underlying causes and symptoms exhibits a distinct set of dominant modules, and their distribution consistently aligns with those modules’ functionalities (finding 8 and 10). Leveraging these insights, researchers can target bug detection and repair efforts toward the most relevant modules, thereby enhancing and optimizing the analysis efficiency.

\subsection{Threats to Validity}
\label{threats}
Our study involves manual analysis, which may introduce subjective bias. To mitigate this threat, we invite 6 domain researchers with at least two years of experience to conduct manual analysis and labeling, ensuring sufficient knowledge of robotics software. This step helped reduce potential misunderstandings. 
Furthermore, the researchers strictly investigate bug reports according to the criteria of reproducibility and adverse symptoms, which ensure the validity of our collected EAIR bugs. In addition, in RQ analysis, two researchers inspected the bugs independently, cross-validated their observations, and discussed them with the other authors to reach a consensus, which mitigates human errors effectively.

%本研究可能受到以下几方面的有效性威胁（threats to validity）：

%人工分析的主观性（Bias in Manual Analysis）
%研究过程中涉及人工分析，可能引入一定的主观偏差。为降低该风险，首先，我们的研究人员在进行人工分析和标注前对传统机器人软件和AI agent相关的知识进行了专业化学习，减少引入认为误解，其次，研究人员严格按照可复现和不良症状两个条件筛选bug repot，确保收集的EAIR bug 真实有效。其次。在第3、4、5节的研究过程中，两位共同作者独立检查缺陷，随后进行交叉验证，并与其他作者讨论以达成共识，从而最大程度减少人为误差，提高分析的客观性和一致性。

\section{Related Work}
\label{relatedwork}

\para{Bug analysis for robot system}
Existing studies~\cite{kim2019rvfuzzer,kim2021pgfuzz,han2022control,kim2023patchverif, kim2022pgpatch, castellanos2021attkfinder, zhang2019towards, tychalas2021icsfuzz, rajput2023icspatch, wen2020plug,kim2022drivefuzz, huai2023doppelganger} have focused on analyzing software bugs in other types of robotic systems.
In robotic vehicles, researchers primarily utilize testing techniques to identify various software bugs such as configuration bugs~\cite{jung2021swarmbug}, logic flaws~\cite{jung2022swarmflawfinder}, type errors~\cite{kate2021physframe}, correctness bugs~\cite{kim2022robofuzz}, cyber-physical inconsistencies~\cite{choi2020cyber} and so on.
In industrial control systems, program analysis methods are widely employed to inspect system code, focusing on detecting PLC intrusions~\cite{yang2022detecting}, behavioral integrity violations~\cite{adepu2020control}, insecure code patterns~\cite{pogliani2020detecting}, misconfiguration vulnerabilities~\cite{zhang2022automated} and so on.
For autonomous driving, empirical studies have been conducted to reveal the characteristics of system bugs~\cite{jiangleveraging, garcia2020comprehensive}. Further, researchers have also applied various testing techniques to detect various software bugs, including safety violations~\cite{li2020av}, denial-of-service vulnerabilities~\cite{hu2021automated}, driving bugs~\cite{kim2022drivefuzz}, semantic DoS attacks~\cite{wan2022too}, adversarial driving maneuvers~\cite{song2023discovering} and so on.
However, none of these studies can support a systematic examination of EAIR system bugs.
%since EAIR systems feature architectures (see Figure~\ref{EAIR_architecture}) and mechanisms (see Section~\ref{EAIRs}) that are fundamentally different from these other types of robotic systems, existing bug analysis techniques are not readily applicable to EAIR software. %Moreover, there is currently a lack of research specifically dedicated to analyzing software bugs in EAIR systems.

%当前已经存在一系列研究聚焦在分析其他领域机器人的bug。 这些研究通常可以被分为以下三类(1)robotic vehicles [xx] (2) industrial control systems, and (3) autonomous driving. 
%在robotic vehicles 方面，研究人员主要通过模糊测试等测试分析方法探索了各种系统软件bug，包括配置错误，逻辑缺陷， GPS 欺骗攻击，类型错误，正确性错误，ROS漏洞，输入验证错误，网络物理不一致漏洞，范围规范错误等。
%而在industrial control systems，研究人员通过程序分析方法针对系统代码进行审查，包括攻击媒介检测， PLC 入侵，行为完整性检测，不安全代码模式识别，错误配置漏洞识别等。
%，此外，在自动驾驶方面，研究人员开展实证研究发现系统bug特征[xx]，利用模糊测试等测试分析技术检测自动驾驶软件中的各种bug，包括安全违规行为，拒绝服务漏洞，驾驶错误，语义 DoS 漏洞，对抗性驾驶动作。然而，由于EAIR系统软件具有与这些机器人软件截然不同的架构（图4）,以及特有机制（Section 2），这些bug分析无法适用于EAIR的bug分析。此外，当前仍然缺乏EAIR系统软件bug分析相关的研究

\para{EAIR system robustness}
%In the EAIR domain, ensuring the robustness of software systems is significant, as any software defect can pose serious risks to users and EAIR itself. 
To the best of our knowledge, the most relevant research on EAIR system robustness is a limited number of survey studies~\cite{duan2022survey, liu2024aligning,roy2021machine,xu2024survey}. Duan et al. conducted a comprehensive survey of embodied simulators, identifying potential challenges in their implementation~\cite{duan2022survey}. Other studies performed a systematic investigation of embodied intelligent robotic systems, summarizing the implementation problems and challenges encountered in this wild~\cite{liu2024aligning,roy2021machine,xu2024survey}. These works are incapable of analyzing or mitigating EAIR bugs. 
%In contrast, our study marks the first attempt to systematically collect, categorize, and characterize EAIR software bugs. This work represents a critical first step toward systematically eliminating EAIR software bugs and provides practical insights for future research on bug detection and repair.

%在EAIR领域，确保软件系统的稳健性（robustness）是首要任务，因为任何软件缺陷都可能对用户及EAIR本身造成严重损害。然而，当前与EAIR system robustness相关的是几个实证工作[xx]。
%Duan等人对具身智能的仿真器进行了全面的调查，并总结了具身智能的仿真器潜在的实现问题和挑战。Liu等人对具身智能机器人系统进行了系统性的调查，总结具身智能的机器人遇到的实现问题和挑战。这些工作都不能有效地分析和避免EAIR bug，本研究首次系统性地收集、分类并刻画自动驾驶软件缺陷。这一研究是迈向系统性消除EAIR软件bug的关键第一步，为未来针对自动驾驶系统软件缺陷的检测、修复及稳健性提升提供了重要的理论与实践基础。

\section{Conclusion}
\label{Conclusion}

In this study, we first systematically characterize and taxonomize bugs in EAIR software by analyzing a total of 885 software bugs collected from 80 real-world EAIR software projects. From this analysis, we identified 18 underlying causes, 15 distinct symptoms, and 13 affected system components.
For developers, we summarized 10 key findings aimed at assisting them in more effectively identifying and repairing EAIR software bugs.
For researchers, we reveal a set of critical challenges from our results, highlighting areas that necessitate further in-depth research to advance the bug testing and repair for EAIR systems.
Furthermore, our findings lay a solid foundation for future research aimed at enhancing the quality assurance practices of EAIR software.

%\section*{Data Availability}
%To facilitate future research, we will release our empirical study results as well as the corresponding datasets in a public repository (\url{https://doi.org/10.6084/m9.figshare.28631276.v1}). 

%在本研究中，我们系统性地刻画并分类（characterized and taxonomized）了EAIR软件中的缺陷，分析了来自80个真实的EAIR软件项目共885个软件bug，并归纳出其18种根本原因（underlying causes）、15种缺陷表现（symptoms）以及13个受影响的系统组件。
%本研究的成果可为AV系统的研究人员和开发者提供重要指导：
%对于开发者，我们总结了11项关键研究发现，帮助他们更高效地识别和修复软件缺陷。
%对于研究人员，我们从研究结果中提炼了多个关键挑战，呼吁开展进一步深入研究，以推进AV软件测试与缺陷修复技术的发展。
%此外，本研究的发现还为未来AV软件质量保障技术的设计与优化奠定了基础。

%% file: tables/symptoms.tex
\begin{table}[b]

\footnotesize
\caption{Distribution of bug symptoms in EAIR software.}
\centering

\begin{tabular}{|lll|} 
\hline
\rowcolor[rgb]{0.753,0.753,0.753} \textbf{Symptom} & \textbf{\#Bug} & \textbf{Ratio}  \\ 
\hline
\multicolumn{3}{|c|}{EAIR-specific Symptoms}                                          \\ 
\hline
S1. Insufficient reliability of AI agent output (RAIO)        & 104            & 11.05\%         \\
S2. Unauthorized access and execution of AI agent       & 4              & 0.43\%          \\
S3. Training failures (TF)                                  & 26             & 2.76\%          \\
S4. Conflicts and collisions (CC)                           & 42             & 4.46\%          \\
S5. 3D environment modeling failures (3DEMF)                  & 20             & 2.13\%          \\
S6. Poor compatibility of agent (PCA)                       & 23             & 2.44\%          \\
S7. Violation on physical reality (VPR)                     & 40             & 4.25\%          \\
S8. Embodied control anomaly (ECA)                          & 89             & 9.46\%         \\ 
\hline
\multicolumn{3}{|c|}{General System Symptoms}                                       \\ 
\hline
S9. Crashes                                            & 170            & 18.07\%         \\
S10. Hangs                                              & 88             & 9.35\%          \\
S11. Build errors (Build)                                      & 57             & 6.06\%          \\
S12. Display and GUI errors (GUI)                            & 136            & 14.45\%         \\
S13. Performance degradation (PD)                            & 17             & 1.81\%          \\
S14. Compilation failures (Comp)                              & 21             & 2.23\%          \\
S15. Hardware interaction failures (HIF)                     & 3              & 0.32\%          \\
\hline
%Unknown                                            & 101             & 10.73\%         \\
%\hline
\end{tabular}

\label{symptom}
\end{table}

%% file: tables/EAIR_bug_taxonomy.tex
\begin{table}[]
\footnotesize
\centering
\caption{The taxonomy of EAIR system bug.}

\begin{tabular}{l|l|l}
\hline
\rowcolor[HTML]{9B9B9B} 
\multicolumn{1}{c|}{\cellcolor[HTML]{9B9B9B}\textbf{Category}}                    & \multicolumn{1}{c|}{\cellcolor[HTML]{9B9B9B}\textbf{Type}}  & \multicolumn{1}{c|}{\cellcolor[HTML]{9B9B9B}\textbf{Ratio}}                                                                                                                                                                                                                                             \\ \hline
%\multicolumn{1}{c|}{}                                                             & \multicolumn{1}{c}{Bug}                                                                                     & \multicolumn{1}{c|}{Ratio}                  &   \multicolumn{1}{c|}{ Cause (\#Cause) }                                                                                                                                                                                                             \\ \hline
                                                                                  & \cellcolor[HTML]{EFEFEF}Perception and reasoning error                                                      & \cellcolor[HTML]{EFEFEF}12.88\% (114/885)  \\
\multirow{-2}{*}{\begin{tabular}[c]{@{}l@{}}Embodied \\ perception\end{tabular}}  & 3D scene construction error                                                                                 & 2.82\% (25/885)                                                                                                                                                                                                                                  \\ \hline
                                                                                  & \cellcolor[HTML]{EFEFEF}EQA hallucination                                                                   & \cellcolor[HTML]{EFEFEF}0.90\% (8/885)                                                                                                                                                                                   \\
\multirow{-2}{*}{\begin{tabular}[c]{@{}l@{}}Embodied \\ interaction\end{tabular}} & Embodied grasping failure                                                                                   & 12.99\% (115/885)                                                                \\ \hline
                                                                                  & \cellcolor[HTML]{EFEFEF}Machine learning issue                                                              & \cellcolor[HTML]{EFEFEF}15.71\% (139/885)                       \\
\multirow{-2}{*}{\begin{tabular}[c]{@{}l@{}}Embodied \\ control\end{tabular}}     & Navigation semantic error                                                                                   & 2.94\% (26/885)                                                                                                                                                                             \\ \hline
                                                                                  & \cellcolor[HTML]{EFEFEF}\begin{tabular}[c]{@{}l@{}}Physical-feature modeling \\ incompleteness\end{tabular} & \cellcolor[HTML]{EFEFEF}18.98\% (168/885)    \\
                                                                                  & Simulation enviorment error                                                                                 & 0.90\% (8/885)                                                                                                                                                                                                                                   \\
                                                                                  & \cellcolor[HTML]{EFEFEF}Object interaction Inconsistency                                                    & \cellcolor[HTML]{EFEFEF}1.36\% (12/885)                                                                                                                                                                                   \\
\multirow{-4}{*}{\begin{tabular}[c]{@{}l@{}}Embodied \\ simulation\end{tabular}} & Agent conflicts and collisions                                                                              & 4.75\% (42/885)                                                                                                                                                                                     \\ \hline
\rowcolor[HTML]{EFEFEF} 
Other                                                                             &  -                                                                                       & 25.76\% (228/885)                                                                                                    \\ \hline
\end{tabular}
\label{taxonomy}
\end{table}

%% file: tables/RootCause_copy.tex
\begin{table*}[t]
    \footnotesize
    \caption{Cross-tabulation of EAIR bug categories and root causes.}
    \centering
    \setlength{\tabcolsep}{1.1mm}
\begin{tabular}{l|cccccccc|cccccccccc}
    \hline
    & \multicolumn{8}{c|}{\textbf{EAIR-specific Causes}} & \multicolumn{10}{c}{\textbf{Traditional Causes}} \\ 
    \diagbox[width=3.9cm]{Root cause}{Bug} & \rotatebox{90}{\textbf{IAIO}} & \rotatebox{90}{\textbf{PLAI}} & \rotatebox{90}{\textbf{AAI}} & \rotatebox{90}{\textbf{AIMD}} & \rotatebox{90}{\textbf{MIAE}} & \rotatebox{90}{\textbf{SRI}} & \rotatebox{90}{\textbf{ESME}} & \rotatebox{90}{\textbf{EMII}} & \rotatebox{90}{\textbf{INC}} & \rotatebox{90}{\textbf{IA}} & \rotatebox{90}{\textbf{ACI}} & \rotatebox{90}{\textbf{IDS}} & \rotatebox{90}{\textbf{MAPI}} & \rotatebox{90}{\textbf{Con}} & \rotatebox{90}{\textbf{MEM}} & \rotatebox{90}{\textbf{ID}} & \rotatebox{90}{\textbf{CCE}} & \rotatebox{90}{\textbf{Other}} \\ 
    \hline
    \rowcolor[HTML]{EFEFEF}
    Perception and reasoning error & 2 & 2 & 2 & 3 & 1 & 0 & 0 & 4 & 8 & 30 & 25 & 9 & 15 & 3 & 2 & 5 & 0 & 3 \\
    3D scene construction error & 0 & 0 & 6 & 0 & 3 & 0 & 0 & 0 & 0 & 2 & 3 & 7 & 1 & 0 & 0 & 2 & 0 & 1 \\
    \rowcolor[HTML]{EFEFEF}
    EQA hallucination & 0 & 0 & 0 & 0 & 0 & 0 & 0 & 0 & 1 & 3 & 3 & 0 & 1 & 0 & 0 & 0 & 0 & 0 \\
    Embodied grasping failure & 2 & 1 & 1 & 2 & 2 & 3 & 0 & 2 & 12 & 22 & 28 & 6 & 24 & 7 & 3 & 0 & 0 & 0 \\
    \rowcolor[HTML]{EFEFEF}
    Machine learning issue & 3 & 0 & 1 & 20 & 0 & 0 & 0 & 4 & 13 & 25 & 21 & 14 & 21 & 1 & 6 & 8 & 0 & 2 \\
    Navigation semantic error & 2 & 0 & 0 & 0 & 2 & 0 & 0 & 0 & 2 & 6 & 6 & 0 & 6 & 1 & 0 & 1 & 0 & 0 \\
    \rowcolor[HTML]{EFEFEF}
    \begin{tabular}[c]{@{}l@{}}Physical-feature modeling \\ incompleteness\end{tabular} & 5 & 1 & 0 & 0 & 10 & 0 & 2 & 3 & 17 & 36 & 44 & 13 & 21 & 1 & 7 & 5 & 2 & 1 \\
    Simulation environment error & 0 & 0 & 0 & 6 & 0 & 0 & 2 & 0 & 0 & 0 & 0 & 0 & 0 & 0 & 0 & 0 & 0 & 0 \\
    \rowcolor[HTML]{EFEFEF}
    Object interaction Inconsistency & 0 & 0 & 0 & 0 & 0 & 8 & 0 & 0 & 0 & 0 & 0 & 0 & 4 & 0 & 0 & 0 & 0 & 0 \\
    Agent conflicts and collisions & 4 & 0 & 0 & 0 & 0 & 0 & 0 & 0 & 17 & 10 & 5 & 6 & 0 & 0 & 0 & 0 & 0 & 0 \\
    \rowcolor[HTML]{EFEFEF}
    Other & 1 & 3 & 0 & 0 & 2 & 1 & 0 & 0 & 7 & 40 & 30 & 19 & 31 & 4 & 5 & 71 & 1 & 11 \\
    \hline
    Total & 19 & 7 & 10 & 31 & 20 & 12 & 4 & 11 & 77 & 174 & 165 & 74 & 124 & 17 & 23 & 92 & 3 & 18 \\
    \rowcolor[HTML]{EFEFEF}
    Ratio (\%) & 2.15 & 0.79 & 1.13 & 3.50 & 2.26 & 1.36 & 0.45 & 1.24 & 8.70 & 19.66 & 18.64 & 8.36 & 14.01 & 1.92 & 2.60 & 10.40 & 0.34 & 2.03 \\ \hline
    \end{tabular} 
    \label{crosstable_transposed_updated}
\end{table*}

%% file: tables/causeandsymptom.tex
\begin{table*}[t]

  \footnotesize
  \caption{Frequency of symptoms that each root cause may produce.}
  \centering
  \setlength{\tabcolsep}{2.2mm}
  
  \begin{tabular}{
  >{\columncolor[HTML]{C0C0C0}}l|
  >{\columncolor[HTML]{C0C0C0}}c
  >{\columncolor[HTML]{C0C0C0}}c
  >{\columncolor[HTML]{C0C0C0}}c
  >{\columncolor[HTML]{C0C0C0}}c
  >{\columncolor[HTML]{C0C0C0}}c
  >{\columncolor[HTML]{C0C0C0}}c
  >{\columncolor[HTML]{C0C0C0}}c
  >{\columncolor[HTML]{C0C0C0}}c
  >{\columncolor[HTML]{C0C0C0}}c
  >{\columncolor[HTML]{C0C0C0}}c
  >{\columncolor[HTML]{C0C0C0}}c
  >{\columncolor[HTML]{C0C0C0}}c
  >{\columncolor[HTML]{C0C0C0}}c
  >{\columncolor[HTML]{C0C0C0}}c
  >{\columncolor[HTML]{C0C0C0}}c}
  %>{\columncolor[HTML]{C0C0C0}}c}
  %>{\columncolor[HTML]{C0C0C0}}c }
  \hline
  \rowcolor[HTML]{FFFFFF}
    \diagbox[width=2.4cm]{Root cause}{Symptom} & RAIO & UAEAI & TF & CC & 3DEMF & PCA & VPR & ECA & Crash & Hang & Build & GUI & PD & Comp & HIF  \\ \hline
  \rowcolor[HTML]{EFEFEF}
  IAIO                             & 1    & 0     & 1    & 3    & 0     & 1     & 0    & 1     & 3     & 1     & 0     & 8    & 0    & 0     & 0          \\
  \rowcolor[HTML]{FFFFFF}
  PLAI                            & 0    & 4     & 0    & 0    & 0     & 0     & 0    & 0     & 1     & 1     & 1     & 0    & 0    & 0     & 0          \\
  \rowcolor[HTML]{EFEFEF}
  AAI                              & 6    & 0     & 1    & 0    & 1     & 0     & 0    & 0     & 2     & 0     & 0     & 0    & 0    & 0     & 0          \\
  \rowcolor[HTML]{FFFFFF}
  AIMD                             & 16   & 0     & 5    & 1    & 0     & 0     & 1    & 1     & 3     & 1     & 1     & 1    & 0    & 0     & 0         \\
  \rowcolor[HTML]{EFEFEF}
  MIAE                             & 0    & 0     & 0    & 1    & 2     & 7     & 4    & 0     & 3     & 3     & 0     & 0    & 0    & 0     & 2          \\
  \rowcolor[HTML]{FFFFFF}
  SRI                              & 2    & 0     & 0    & 1    & 1     & 0     & 5    & 0     & 0     & 0     & 0     & 0    & 0    & 0     & 0         \\
  \rowcolor[HTML]{EFEFEF}
  ESME                             & 0    & 0     & 0    & 1    & 0     & 1     & 5    & 0     & 1     & 1     & 1     & 0    & 1    & 0     & 0         \\
  \rowcolor[HTML]{FFFFFF}
  EMII                             & 0    & 0     & 4    & 0    & 5     & 0     & 1    & 0     & 1     & 0     & 0     & 0    & 0    & 0     & 1          \\
  \rowcolor[HTML]{EFEFEF}
  INC                              & 11   & 0     & 1    & 6    & 0     & 0     & 0    & 7     & 10    & 3     & 3     & 31   & 1    & 0     & 0          \\
  \rowcolor[HTML]{FFFFFF}
  IA                               & 28   & 0     & 5    & 16   & 4     & 5     & 7    & 27    & 17    & 12    & 12    & 41   & 3    & 2     & 0          \\
  \rowcolor[HTML]{EFEFEF}
  ACI                              & 18   & 0     & 2    & 6    & 3     & 0     & 6    & 23    & 40    & 24    & 6     & 29   & 6    & 3     & 0          \\
  \rowcolor[HTML]{FFFFFF}
  IDS                              & 8    & 0     & 3    & 1    & 0     & 2     & 3    & 1     & 26    & 12    & 6     & 8    & 3    & 1     & 0          \\
  \rowcolor[HTML]{EFEFEF}
  MAPI                             & 14   & 0     & 1    & 3    & 4     & 5     & 5    & 27    & 39    & 10    & 7     & 10   & 2    & 10    & 0          \\
  \rowcolor[HTML]{FFFFFF}
  Con                              & 0    & 0     & 0    & 0    & 0     & 0     & 0    & 2     & 7     & 10    & 0     & 0    & 1    & 0     & 0          \\
  \rowcolor[HTML]{EFEFEF}
  MEM                              & 0    & 0     & 2    & 2    & 0     & 2     & 1    & 0     & 12    & 1     & 1     & 2    & 0    & 0     & 0          \\
  \rowcolor[HTML]{FFFFFF}
  ID                               & 0    & 0     & 0    & 0    & 0     & 0     & 0    & 0     & 3     & 6     & 19    & 6    & 0    & 5     & 0         \\
  \rowcolor[HTML]{EFEFEF}
  CCE                              & 0    & 0     & 0    & 1    & 0     & 0     & 1    & 0     & 1     & 1     & 0     & 0    & 0    & 0     & 0          \\
  \rowcolor[HTML]{FFFFFF}
  Other                            & 0    & 0     & 1    & 0    & 0     & 0     & 1    & 0     & 1     & 2     & 0     & 0    & 0    & 0     & 0         \\ \hline
  \rowcolor[HTML]{EFEFEF}
  Total                            & 104  & 4     & 26   & 42   & 20    & 23    & 40   & 89    & 170   & 88    & 57    & 136  & 17   & 21    & 3        \\ \hline
  \end{tabular}
  
  \label{causeandsymptom}
  \end{table*}

%% file: tables/compositeBug.tex
\begin{table}[b]

\footnotesize
\caption{Distribution of compound bugs in EAIR software.}
\centering

  \setlength{\tabcolsep}{4mm} {

\begin{tabular}{|l|cc|c|}
\hline
\rowcolor[HTML]{C0C0C0} 
                       & \multicolumn{2}{c|}{\cellcolor[HTML]{C0C0C0}Compound bug} & Non-compound bug      \\ \hline
                       & T1                          & T2                          & T3                    \\ \hline
EAIR-specific bug      & 47 & 32  &    39             \\ \hline
\end{tabular} }

\label{compound}
\end{table}

%% file: tables/causeandmodulenew.tex
\begin{table*}[]

    \footnotesize
    \caption{Distribution of root causes on each core module.}
    \centering
    \setlength{\tabcolsep}{0.3mm}{

\begin{tabular}{l|ccc|cc|ccc|c|c|c|c|c|c|c}
\hline
\rowcolor[HTML]{FFFFFF} 
Module                                             & \multicolumn{3}{c|}{\cellcolor[HTML]{FFFFFF}Embodied perception} & \multicolumn{2}{l|}{\cellcolor[HTML]{FFFFFF}Embodied interaction} & \multicolumn{3}{c|}{\cellcolor[HTML]{FFFFFF}Embodied  control} & \cellcolor[HTML]{FFFFFF}                                                                                & \multicolumn{1}{l|}{\cellcolor[HTML]{FFFFFF}}                           & \multicolumn{1}{l|}{\cellcolor[HTML]{FFFFFF}}                            & \multicolumn{1}{l|}{\cellcolor[HTML]{FFFFFF}}                          & \multicolumn{1}{l|}{\cellcolor[HTML]{FFFFFF}}                          & \multicolumn{1}{l|}{\cellcolor[HTML]{FFFFFF}}                        & \multicolumn{1}{l}{\cellcolor[HTML]{FFFFFF}}                        \\
\rowcolor[HTML]{FFFFFF} 
\diagbox{Root cause}{Submodule} & Reason              & Sensors              & 3Dscene             & EQA                           & Grasping                          & Learning            & Remote            & Navigation           & \multirow{-2}{*}{\cellcolor[HTML]{FFFFFF}\begin{tabular}[c]{@{}c@{}}Embodied\\ simulation\end{tabular}} & \multicolumn{1}{l|}{\multirow{-2}{*}{\cellcolor[HTML]{FFFFFF}Executor}} & \multicolumn{1}{l|}{\multirow{-2}{*}{\cellcolor[HTML]{FFFFFF}Interface}} & \multicolumn{1}{l|}{\multirow{-2}{*}{\cellcolor[HTML]{FFFFFF}Config.}} & \multicolumn{1}{l|}{\multirow{-2}{*}{\cellcolor[HTML]{FFFFFF}Display}} & \multicolumn{1}{l|}{\multirow{-2}{*}{\cellcolor[HTML]{FFFFFF}Other}} & \multicolumn{1}{l}{\multirow{-2}{*}{\cellcolor[HTML]{FFFFFF}Total}} \\ \hline
\rowcolor[HTML]{EFEFEF} 
IAIO                                               & 1                   & 1                    & 0                   & 0                             & 2                                 & 3                   & 0                 & 2                    & 9                                                                                                       & 0                                                                       & 0                                                                        & 0                                                                      & 1                                                                      & \textbf{0}                                                           & 19                                                                  \\
\rowcolor[HTML]{FFFFFF} 
PLAI                                              & 0                   & 2                    & 0                   & 0                             & 1                                 & 0                   & 0                 & 0                    & 1                                                                                                       & 0                                                                       & 0                                                                        & 3                                                                      & 0                                                                      & 0                                                                    & 7                                                                   \\
\rowcolor[HTML]{EFEFEF} 
AAI                                                & 2                   & 0                    & 6                   & 0                             & 1                                 & 1                   & 0                 & 0                    & 0                                                                                                       & 0                                                                       & 0                                                                        & 0                                                                      & 0                                                                      & 0                                                                    & 10                                                                  \\
\rowcolor[HTML]{FFFFFF} 
AIMD                                               & 3                   & 0                    & 0                   & 0                             & 2                                 & 20                  & 0                 & 0                    & 6                                                                                                       & 0                                                                       & 0                                                                        & 0                                                                      & 0                                                                      & 0                                                                    & 31                                                                  \\
\rowcolor[HTML]{EFEFEF} 
MIAE                                               & 1                   & 0                    & 3                   & 0                             & 2                                 & 0                   & 1                 & 1                    & 10                                                                                                      & 0                                                                       & 0                                                                        & 0                                                                      & 0                                                                      & 2                                                                    & 20                                                                  \\
\rowcolor[HTML]{FFFFFF} 
SRI                                                & 0                   & 0                    & 0                   & 0                             & 3                                 & 0                   & 0                 & 0                    & 8                                                                                                       & 0                                                                       & 0                                                                        & 0                                                                      & 0                                                                      & 1                                                                    & 12                                                                  \\
\rowcolor[HTML]{EFEFEF} 
ESME                                               & 0                   & 0                    & 0                   & 0                             & 2                                 & 0                   & 0                 & 0                    & 4                                                                                                       & 0                                                                       & 0                                                                        & 1                                                                      & 0                                                                      & 1                                                                    & 8                                                                   \\
\rowcolor[HTML]{FFFFFF} 
EMII                                               & 0                   & 4                    & 0                   & 0                             & 0                                 & 4                   & 0                 & 0                    & 3                                                                                                       & 0                                                                       & 0                                                                        & 0                                                                      & 0                                                                      & 0                                                                    & 11                                                                  \\
\rowcolor[HTML]{EFEFEF} 
INC                                                & 3                   & 5                    & 0                   & 1                             & 12                                & 13                  & 0                 & 2                    & 34                                                                                                      & 4                                                                       & 1                                                                        & 1                                                                      & 1                                                                      & 0                                                                    & 77                                                                  \\
\rowcolor[HTML]{FFFFFF} 
IA                                                 & 8                   & 22                   & 2                   & 3                             & 22                                & 25                  & 2                 & 4                    & 46                                                                                                      & 5                                                                       & 5                                                                        & 17                                                                     & 0                                                                      & 13                                                                   & 174                                                                 \\
\rowcolor[HTML]{EFEFEF} 
ACI                                                & 15                  & 10                   & 3                   & 3                             & 28                                & 21                  & 1                 & 5                    & 49                                                                                                      & 5                                                                       & 10                                                                       & 7                                                                      & 0                                                                      & 8                                                                    & 165                                                                 \\
\rowcolor[HTML]{FFFFFF} 
IDS                                                & 1                   & 8                    & 7                   & 0                             & 6                                 & 14                  & 0                 & 0                    & 19                                                                                                      & 6                                                                       & 1                                                                        & 7                                                                      & 0                                                                      & 5                                                                    & 74                                                                  \\
\rowcolor[HTML]{EFEFEF} 
MAPI                                               & 9                   & 6                    & 1                   & 1                             & 24                                & 21                  & 4                 & 2                    & 25                                                                                                      & 5                                                                       & 5                                                                        & 19                                                                     & 1                                                                      & 1                                                                    & 124                                                                 \\
\rowcolor[HTML]{FFFFFF} 
Con                                                & 0                   & 3                    & 0                   & 0                             & 7                                 & 1                   & 0                 & 1                    & 1                                                                                                       & 1                                                                       & 3                                                                        & 0                                                                      & 0                                                                      & 0                                                                    & 17                                                                  \\
\rowcolor[HTML]{EFEFEF} 
MEM                                                & 1                   & 1                    & 0                   & 0                             & 3                                 & 6                   & 0                 & 0                    & 7                                                                                                       & 1                                                                       & 0                                                                        & 3                                                                      & 0                                                                      & 1                                                                    & 23                                                                  \\
\rowcolor[HTML]{FFFFFF} 
ID                                                 & 1                   & 4                    & 2                   & 0                             & 0                                 & 8                   & 1                 & 0                    & 5                                                                                                       & 2                                                                       & 3                                                                        & 17                                                                     & 2                                                                      & 47                                                                   & 92                                                                  \\
\rowcolor[HTML]{EFEFEF} 
CCE                                                & 0                   & 0                    & 0                   & 0                             & 0                                 & 0                   & 0                 & 0                    & 2                                                                                                       & 0                                                                       & 1                                                                        & 0                                                                      & 0                                                                      & 0                                                                    & 3                                                                   \\
\rowcolor[HTML]{FFFFFF} 
Other                                              & 0                   & 3                    & 1                   & 0                             & 0                                 & 2                   & 0                 & 0                    & 1                                                                                                       & 0                                                                       & 1                                                                        & 0                                                                      & 0                                                                      & 10                                                                   & 18                                                                  \\ \hline
\end{tabular}    
    }

    \label{causeandmodule}
    \end{table*}

%% file: tables/symptomandmodulenew.tex
\begin{table*}[t]

    \footnotesize
    \caption{Distribution of symptoms on each core module.}
    \centering
    \setlength{\tabcolsep}{0.3mm}{

    \begin{tabular}{l|ccc|cc|ccc|c|c|c|c|c|c|c}
\hline
Module                                          & \multicolumn{3}{c|}{Embodied perception} & \multicolumn{2}{c|}{Embodied interaction} & \multicolumn{3}{c|}{Embodied control} & \multicolumn{1}{l|}{}                                                                                & \multicolumn{1}{l|}{}                           & \multicolumn{1}{l|}{}                            & \multicolumn{1}{l|}{}                          & \multicolumn{1}{l|}{}                          & \multicolumn{1}{l|}{}                        & \multicolumn{1}{l}{}                        \\
\diagbox{Symptom}{Submodule} & Sensors      & Reason      & 3DScene     & Grasping               & EQA              & Learning    & Navigation   & Remote   & \multicolumn{1}{l|}{\multirow{-2}{*}{\begin{tabular}[c]{@{}l@{}}Embodied\\ simulation\end{tabular}}} & \multicolumn{1}{l|}{\multirow{-2}{*}{Executor}} & \multicolumn{1}{l|}{\multirow{-2}{*}{Interface}} & \multicolumn{1}{l|}{\multirow{-2}{*}{Config.}} & \multicolumn{1}{l|}{\multirow{-2}{*}{Display}} & \multicolumn{1}{l|}{\multirow{-2}{*}{Other}} & \multicolumn{1}{l}{\multirow{-2}{*}{Total}} \\ \hline
\rowcolor[HTML]{EFEFEF} 
RAIO                                            & 2            & 12          & 5           & 3                      & 1                & 55          & 1            & 0        & 14                                                                                                   & 3                                               & 0                                                & 8                                              & 0                                              & 0                                            & 104                                         \\
UAEAI                                           & 2            & 0           & 0           & 0                      & 0                & 0           & 0            & 0        & 0                                                                                                    & 0                                               & 0                                                & 2                                              & 0                                              & 0                                            & 4                                           \\
\rowcolor[HTML]{EFEFEF} 
TF                                              & 1            & 0           & 0           & 1                      & 0                & 20          & 0            & 0        & 2                                                                                                    & 0                                               & 1                                                & 0                                              & 0                                              & 1                                            & 26                                          \\
CC                                              & 1            & 2           & 0           & 12                     & 0                & 1           & 3            & 0        & 21                                                                                                   & 1                                               & 1                                                & 0                                              & 0                                              & 0                                            & 42                                          \\
\rowcolor[HTML]{EFEFEF} 
3DEMF                                           & 10           & 0           & 4           & 0                      & 0                & 0           & 0            & 0        & 5                                                                                                    & 0                                               & 0                                                & 1                                              & 0                                              & 0                                            & 20                                          \\
PCA                                             & 3            & 1           & 0           & 0                      & 0                & 3           & 0            & 0        & 9                                                                                                    & 2                                               & 0                                                & 2                                              & 0                                              & 3                                            & 23                                          \\
\rowcolor[HTML]{EFEFEF} 
VPR                                             & 4            & 0           & 2           & 11                     & 0                & 3           & 0            & 0        & 18                                                                                                   & 1                                               & 0                                                & 1                                              & 0                                              & 0                                            & 40                                          \\
ECA                                             & 8            & 6           & 1           & 45                     & 2                & 1           & 3            & 5        & 10                                                                                                   & 4                                               & 1                                                & 2                                              & 0                                              & 1                                            & 89                                          \\
\rowcolor[HTML]{EFEFEF} 
Crash                                           & 4            & 19          & 7           & 23                     & 1                & 33          & 4            & 2        & 40                                                                                                   & 8                                               & 9                                                & 15                                             & 0                                              & 5                                            & 170                                         \\
Hang                                            & 10           & 4           & 4           & 15                     & 0                & 10          & 2            & 0        & 11                                                                                                   & 7                                               & 8                                                & 5                                              & 1                                              & 11                                           & 88                                          \\
\rowcolor[HTML]{EFEFEF} 
Build                                           & 8            & 0           & 0           & 2                      & 1                & 7           & 0            & 0        & 6                                                                                                    & 0                                               & 4                                                & 20                                             & 2                                              & 7                                            & 57                                          \\
GUI                                             & 3            & 0           & 3           & 6                      & 3                & 4           & 4            & 1        & 94                                                                                                   & 0                                               & 3                                                & 5                                              & 2                                              & 8                                            & 136                                         \\
\rowcolor[HTML]{EFEFEF} 
PD                                              & 0            & 0           & 1           & 1                      & 0                & 3           & 2            & 1        & 3                                                                                                    & 2                                               & 0                                                & 2                                              & 0                                              & 2                                            & 17                                          \\
Comp                                            & 0            & 0           & 0           & 7                      & 0                & 1           & 0            & 0        & 3                                                                                                    & 0                                               & 1                                                & 8                                              & 0                                              & 1                                            & 21                                          \\
\rowcolor[HTML]{EFEFEF} 
HIF                                             & 1            & 0           & 0           & 0                      & 0                & 0           & 1            & 0        & 1                                                                                                    & 0                                               & 0                                                & 0                                              & 0                                              & 0                                            & 3                                           \\
Unknown                                         & 12           & 2           & 1           & 4                      & 0                & 7           & 0            & 0        & 13                                                                                                   & 2                                               & 2                                                & 7                                              & 0                                              & 51                                           & 101                                         \\ \hline
\end{tabular} }

\label{symptomandmodule}
\end{table*}